\documentclass[12pt,nofootinbib,tightenlines,preprintnumbers,superscriptaddress]{revtex4}
\usepackage{graphicx}

\usepackage{amssymb}
\usepackage{amsmath}
\usepackage{color}

\newcommand{\be}{\begin{equation}}
\newcommand{\ee}{\end{equation}}
\newcommand{\ba}{\begin{eqnarray}}
\newcommand{\ea}{\end{eqnarray}}

\newcommand{\bi}{\begin{itemize}}
\newcommand{\ei}{\end{itemize}}
\newcommand{\<}{\langle}
\renewcommand{\>}{\rangle}

\newcommand{\la}{\label}

\newcommand{\ahvp}{a_\mu^{\rm hvp}}

\newcommand{\fm}{{\rm{fm}}}

\newcommand{\zv}{Z_{\rm V}}
\newcommand{\Nf}{N_{\rm f}}

\renewcommand{\vec}[1]{\boldsymbol{#1}}

\newcommand{\bv}{b_{\rm V}}

\newcommand{\fv}{f_{\rm V}}

\newcommand{\bbarv}{\overline{b}_{\rm V}}
\newcommand{\mqav}{m_{\rm q}^{\rm av}}
\newcommand{\mql}{m_{{\rm q},l}}
\newcommand{\mqs}{m_{{\rm q},s}}

\newcommand{\MeV}{\mathrm{MeV}}
\newcommand{\GeV}{\mathrm{GeV}}

\begin{document}
\title{The leading hadronic contribution to $(g-2)_\mu$ from lattice QCD with $N_{\rm f}=2+1$ flavours of 
O($a$) improved Wilson quarks}

\preprint{MITP/19-021}
      
\author{Antoine G\'erardin}
\affiliation{John von Neumann Institute for Computing, DESY, Platanenallee 6, D-15738 Zeuthen, Germany}

\author{Marco C\`e} 
\affiliation{Helmholtz-Institut Mainz, Johannes Gutenberg-Universit\"at Mainz,
D-55099 Mainz, Germany}

\author{Georg~von~Hippel}
 \affiliation{PRISMA$^+$ Cluster of Excellence  \& Institut f\"ur Kernphysik,
Johannes Gutenberg-Universit\"at Mainz, D-55099 Mainz, Germany}

\author{Ben H\"orz}
\affiliation{Nuclear Science Division, Lawrence Berkeley National Laboratory, Berkeley, CA 94720, USA}

\author{Harvey B.\ Meyer} 
\affiliation{Helmholtz-Institut Mainz, Johannes Gutenberg-Universit\"at Mainz,
D-55099 Mainz, Germany}
 \affiliation{PRISMA$^+$ Cluster of Excellence  \& Institut f\"ur Kernphysik,
Johannes Gutenberg-Universit\"at Mainz, D-55099 Mainz, Germany}

\author{Daniel Mohler}
\affiliation{Helmholtz-Institut Mainz, Johannes Gutenberg-Universit\"at Mainz,
D-55099 Mainz, Germany}
 \affiliation{PRISMA$^+$ Cluster of Excellence  \& Institut f\"ur Kernphysik,
Johannes Gutenberg-Universit\"at Mainz, D-55099 Mainz, Germany}

\author{Konstantin Ottnad}
 \affiliation{PRISMA$^+$ Cluster of Excellence  \& Institut f\"ur Kernphysik,
Johannes Gutenberg-Universit\"at Mainz, D-55099 Mainz, Germany}

\author{Jonas Wilhelm}
 \affiliation{PRISMA$^+$ Cluster of Excellence  \& Institut f\"ur Kernphysik,
Johannes Gutenberg-Universit\"at Mainz, D-55099 Mainz, Germany}

\author{Hartmut Wittig}
\affiliation{Helmholtz-Institut Mainz, Johannes Gutenberg-Universit\"at Mainz,
D-55099 Mainz, Germany}
 \affiliation{PRISMA$^+$ Cluster of Excellence  \& Institut f\"ur Kernphysik,
Johannes Gutenberg-Universit\"at Mainz, D-55099 Mainz, Germany}

\begin{abstract}
The comparison of the theoretical and experimental determinations of the 
anomalous magnetic moment of the muon $(g-2)_\mu$ constitutes one of the strongest tests of the Standard Model at low energies.
In this article, we compute the leading hadronic contribution to $(g-2)_\mu$ using lattice QCD simulations employing Wilson quarks.
Gauge field ensembles at four different lattice spacings and several values of the pion mass down to its physical value are used.
We apply the O($a$) improvement programme with two discretizations of the vector current 
to better constrain the approach to the continuum limit.
The electromagnetic current correlators are computed in the time-momentum representation.
In addition, we perform auxiliary calculations of the pion form factor at timelike momenta
in order to better constrain the tail of the isovector correlator and to correct its dominant finite-size effect.
For the numerically dominant light-quark contribution, we have rescaled the lepton mass by the pion decay constant computed
on each lattice ensemble. We perform a combined chiral and continuum extrapolation to the
physical point, and our final result is $ \ahvp=(720.0\pm12.4_{\rm stat}\,\pm9.9_{\rm syst})\cdot10^{-10}$.
It contains the contributions of quark-disconnected diagrams, and the systematic error has been enlarged to account 
for the missing isospin-breaking effects.
\end{abstract}

\maketitle

\centerline{\today}

\section{Introduction}

Electrons and muons carry a magnetic moment, which is correctly predicted by Dirac's original theory of the
electron to within a permille of precision. The proportionality factor between the spin and the magnetic
moment of the lepton $\ell$ is parameterized by the gyromagnetic ratio
$g$. In Dirac's theory, $g=2$, and one characterizes the deviation of
$g$ from this reference value by $a_\ell=(g-2)_\ell/2$. Testing the ability
of Quantum Electrodynamics (QED) to correctly predict this precision
observable has played a crucial role in the development of quantum
field theory in general.  Presently, the achieved experimental
precision of 540\,ppb on the measurement of the anomalous magnetic
moment of the muon~\cite{Bennett:2006fi}, $a_\mu$, requires the effects of all three
interactions of the Standard Model (SM) of particle physics to be
included in the theory prediction.  In fact, a tension of about 3.5
standard deviations exists between the SM prediction and the
experimental measurement.  For reviews on the subject, we refer the
reader to~\cite{Jegerlehner:2009ry,Blum:2013xva,Jegerlehner:2017gek}.

Presently, the E989 experiment at Fermilab is performing a new direct
measurement of $a_\mu$~\cite{Grange:2015fou}, and a further experiment using a
different experimental technique is planned at J-PARC~\cite{Mibe:2011zz}.  The final goal
of these experiments is to reduce the uncertainty on $a_\mu$ by a
factor of four.  A reduction of the theory error is thus of paramount
importance, as the first results from the Fermilab experiment are
expected within the next few months.  These  will likely
reach the same precision as the current world average.

On the theory side, the precision of the SM prediction for $a_\mu$ is
completely dominated by hadronic uncertainties. The leading
hadronic contribution enters at second order in the fine-structure
constant $\alpha$ via the vacuum polarization and must be determined
at the few-permille level in order to match the upcoming precision
of the direct measurements of $a_\mu$. In this paper we undertake a
first-principles lattice QCD calculation of this hadronic
contribution (see~\cite{Meyer:2018til} for a recent review of previous
lattice results).  A further hadronic effect, the light-by-light
scattering contribution which enters at third order in the
fine-structure constant, currently contributes at a comparable level
to the theory uncertainty budget and is being addressed
both by dispersive and lattice methods
(see~\cite{Blum:2016lnc,Asmussen:2018oip,Colangelo:2017urn} and references therein).

Our calculation of the hadronic vacuum polarization to the anomalous
magnetic moment of the muon, $\ahvp$, fully includes the effects of
the up, down and strange quarks, while the charm quark (whose
contribution to $\ahvp$ is small) is treated only at the valence
level.  We use ensembles of SU(3) gauge field configurations generated
with an O($a$) improved Wilson quark action as part of the Coordinated
Lattice Simulations (CLS)
initiative~\cite{Bruno:2014jqa,Bali:2016umi}. In particular, the generation
of a physical-mass ensemble~\cite{Mohler:2017wnb} (labelled E250) was largely motivated 
by the goal of improving the lattice determination of $\ahvp$.
We use four different
lattice spacings to control the continuum limit, and the $(u,d,s)$
quark masses are varied at constant average quark mass in order to
perform a chiral interpolation to the physical values of the quark
masses~\cite{Bruno:2014jqa}.  Our calculation is performed at equal up and
down quark masses, and no QED effects are included; however, in the
future both of these isospin-breaking effects will be taken into
account as corrections~\cite{Risch:2017xxe,Risch:2018ozp}.

Lattice QCD, which is formulated in Euclidean space, is well suited
for computing $\ahvp$, since the latter only involves the two-point
function of the hadronic component of the electromagnetic current at
spacelike momenta~\cite{Blum:2002ii}. In this work we employ the
representation of $\ahvp$ as a Euclidean-time integral over the
two-point function in the time-momentum
representation (TMR)~\cite{Bernecker:2011gh}, i.e.\ projected to 
vanishing spatial momentum.  This representation does not require a
parameterization of the vacuum polarization function and has a clear
spectral interpretation in terms of vector hadronic states in the
center-of-mass frame.  The main difficulty in obtaining $\ahvp$ with
good statistical precision is that it probes the TMR
correlator at Euclidean times well beyond 2\,fm, where its relative
precision deteriorates rapidly. Therefore a dedicated treatment of the
tail of the correlator which does not compromise the first-principles
nature of the calculation is needed. Here the spectral representation
of the correlator plays a central role.

An important source of systematic uncertainty is the correction to
$\ahvp$ due to the use of a finite spatial torus. On our lattice
ensembles, this finite-size effect (FSE) mostly stems from the tail of
the isovector component of the TMR correlator.  Thanks to precise
relations~\cite{Luscher:1991cf,Meyer:2011um} between the properties of
the discrete quantum states on the torus and the pion form factor at
timelike momenta, we are able to correct for the dominant part of the
FSE. Finally, the quark-disconnected diagrams, while making only a
few-percent contribution to $\ahvp$, require a dedicated set of
calculations for their evaluation, which demand a large computing-time
investment.

The rest of this paper is organized as follows.  
Section~\ref{sec:metho} describes the methodology followed in our calculation,
including the renormalization and improvement of the TMR correlator
and the treatment of the charm contribution. Section~\ref{sec:results}
presents our lattice data and the extraction of the observable $\ahvp$
on each individual lattice ensemble.  In section~\ref{sec:phys}, the
lattice-spacing and quark-mass dependence of these intermediate
results is fitted in order to arrive at our final result.  Finally, we
compare the latter with phenomenological as well as other recent
lattice determinations in section~\ref{sec:discussion}.

\section{Methodology\label{sec:metho}} 

\subsection{Time-momentum correlators}
We start by providing all relevant relations in the continuum and infinite-volume Euclidean theory.
In the time-momentum representation (TMR), the leading-order hadronic vacuum polarization contribution to $(g-2)_\mu$
is given by the convolution integral
\be\la{eq:TMRamu}
   \ahvp = \left(\frac{\alpha}{\pi}\right)^2\int_0^{\infty}\,
   dt\,\widetilde K(t)G(t), 
\ee
where an analytic expression for the QED kernel function
$\widetilde K(t)$ is given in Appendix~B of Ref.\ \cite{DellaMorte:2017dyu}, and
\be\la{eq:Gx0def}
   G(t)\,\delta_{kl} = -\int d^3x\,\Big\< J_k(t,\vec x)\;J_l(0) \Big\>
\ee
is the spatially summed QCD two-point function of the electromagnetic current
 $\vec J=\frac{2}{3}\bar u\vec\gamma u - \frac{1}{3} \bar d \vec\gamma d -\frac{1}{3} \bar s \vec\gamma s + \frac{2}{3} \bar c \vec\gamma c$.
In isospin-symmetric QCD, we can write
\be\la{eq:Gdecomp}
G(t) = \frac{5}{9} G_l(t) + \frac{1}{9}G_s(t) + \frac{4}{9} G_c(t) + G_{\rm disc}(t),
\ee
where $G_f(t)$ denotes a quark-connected contribution associated with flavour $f$
and $G_{\rm disc}(t)$ is the quark-disconnected contribution.
An alternative decomposition based on the isospin  quantum number $I$ yields
\be\la{eq:GdecompI}
G(t) = G^{I=1}(t) + G^{I=0}(t),\qquad 
G^{I=1}(t) = \frac{1}{2} G_l(t).
\ee
Physically, the latter decomposition is more transparent. In particular, at light pion masses 
the dominant finite-size effects, as well as a logarithmic singularity as $m_\pi\to0$, only concern the isovector
contribution, $a_\mu^{{\rm hvp},I=1}$. Computationally however, the disconnected contributions 
are obtained very differently from the connected ones: they are costly and amount only to a few percent of the total.
Therefore, in our numerical analysis there is an interesting interplay between the two choices of bases to compute $\ahvp$.

With $m_\mu$ the muon mass, the kernel behaves as $\widetilde K(t)\sim \frac{\pi^2}{9}m_\mu^2 t^4$ 
for $t\ll m_\mu^{-1}$ and as $\widetilde K(t)\sim 2\pi^2 t^2$ for $t\gg m_\mu^{-1}$.
Since the lattice data for the correlator $G(t)$ is in lattice units, the muon mass must be known in those units, $am_\mu$.
The knowledge of the lattice spacing in $\GeV^{-1}$ thus plays a crucial role in a 
precision determination of $\ahvp$~\cite{DellaMorte:2017dyu,DellaMorte:2017khn}.
There are then two ways to proceed.
In lattice QCD, where often the physical quark masses are reached only after an extrapolation or interpolation,
$\ahvp$ can either be calculated using the fixed, physical value of $m_\mu=105.66\,\MeV$; or the muon mass 
can be rescaled by a quantity with dimension of mass known experimentally~\cite{Feng:2011zk}.
In our calculation, we have explored both paths. In our final results, 
we adopt the ``rescaling strategy'' for the connected light contribution. 
As a rescaling quantity, we choose the pion decay constant $f_\pi$, 
so that we set\footnote{We use the normalization convention $f_\pi\simeq 92$\,MeV.}
\be
am_\mu = \Big(\frac{m_\mu}{f_\pi}\Big)_{\rm pheno}\cdot (af_\pi)_{\rm lattice} = 1.144\cdot (af_\pi)_{\rm lattice}
\ee
on every lattice ensemble. Our choice is motivated, first, by $f_\pi$
being determined precisely and reliably, both in phenomenology and on the
lattice; and secondly, since $f_\pi$ increases with the pion mass,
this choice has the effect of making the $m_\pi$ dependence of $\ahvp$ weaker.
To intuitively understand  the effect of the rescaling, it is instructive to consider
the calculation of the anomalous magnetic moment of the electron; in this case, 
obtaining  $a_e^{\rm hvp} = ({4\alpha^2}/{3}) m_e^2 \Pi_1$ 
requires computing the time moment $\Pi_1\equiv ({1}/{12})\int_0^\infty dt\;t^4\,G(t)$.
Thus the rescaling simply amounts to computing the dimensionless quantity $f_\pi^2 \Pi_1$,
and converting the result into $a_e^{\rm hvp}$  by using the phenomenological value of $(m_e^2/f_\pi^2)$.

\subsection{Simulation parameters\la{sec:simpar}}

\begin{table}[t!]
        \caption{Parameters of the simulations: $\beta=6/g_0^2$ is the bare gauge coupling, $\kappa_{l,s}$ are the hopping parameters
of the light and strange quarks, $a$ is the lattice spacing and $(L,T)$ are the lattice dimensions in space and time.
Ensembles E250 and B450 have periodic boundary conditions in time, all others have open boundary conditions.
The last column contains the number of gauge configurations used.
Ensembles with an asterisk are not included in the final analysis but are used to control finite-size effects.}        
\vskip 0.1in
\begin{centering}
{\footnotesize
\begin{tabular}{lcl@{\hskip 01em}c@{\hskip 01em}l@{\hskip 01em}l@{\hskip 01em}c@{\hskip 01em}c@{\hskip 01em}c@{\hskip 01em}c@{\hskip 01em}l}
        \hline
        id     &       $\quad\beta\quad$       &       $L^3\times T$   &       $a\,[\fm]$       &    $~~~\kappa_l$          &       $~~~\kappa_s$      &
       $m_{\pi}\,[\MeV]$        &       $m_{K}\,[\MeV]$  &        $m_{\pi}L$     &       $L\,[\fm]$       &       conf.                \\
        \hline
H101    & 3.40  &       $32^3\times96$  & 0.08636       &       0.136760        &       0.136760                & 416(5) & 416(5) & 5.8 & 2.8 & 2000 \\  
H102    &               &       $32^3\times96$  &               &       0.136865        &       0.13654934      & 354(5) & 438(4) & 5.0 & 2.8 & 1900 \\  
H105$^*$        &               &       $32^3\times96$  &               &       0.136970        &       0.13634079      & 284(4) & 460(4) & 3.9 & 2.8 & 2800 \\         
N101    &               &       $48^3\times128$ &               &       0.136970        &       0.13634079      & 282(4) & 460(4) & 5.9 & 4.1 & 1500 \\    
     
C101    &               &       $48^3\times96$  &               &       0.137030        &       0.13622204      & 221(2) & 472(8) & 4.7 & 4.1 & 2600 \\  
\hline
B450            & 3.46  &       $32^3\times64$  & 0.07634       &       0.136890        &       0.136890                & 416(4) & 416(4) & 5.2 & 2.4 & 1500 \\  
S400            &               &       $32^3\times96$  &               &       0.136984        &       0.13670239      & 351(4) & 438(5) & 4.3 & 2.4 & 2800\\          
N401    &               &       $48^3\times128$ &               &       0.137062        &       0.13654808      & 287(4) & 462(5)  & 5.3 & 3.7 & 1100 \\  
\hline
H200$^*$        & 3.55  &       $32^3\times96$  & 0.06426       &       0.137000        &       0.137000                & 419(5) &  419(5) & 4.4 & 2.1 & 2000 \\  
N202    &               &       $48^3\times128$ &               &       0.137000        &       0.137000                & 410(5) &  410(5) & 6.4 & 3.1 & 900 \\  
N203    &               &       $48^3\times128$ &               &       0.137080        &       0.13684028      & 345(4) &  441(5) & 5.4 & 3.1  & 1500 \\  
N200    &               &       $48^3\times128$ &               &       0.137140        &       0.13672086      & 282(3) &  463(5) & 4.4 & 3.1  & 1700 \\  
D200    &               &       $64^3\times128$ &               &       0.137200        &       0.13660175      & 200(2) &  480(5) & 4.2 & 4.1 & 2000 \\  
E250    &               &       $96^3\times192$ &               &       0.137233        &       0.13653663      & 130(1) &            & 4.1 & 6.2 & 500 \\  
\hline
N300    & 3.70  &       $48^3\times128$ &0.04981        &       0.137000        &       0.137000                & 421(4) & 421(4) & 5.1 & 2.4 & 1700 \\  
N302    &               &       $48^3\times128$ &               &       0.137064        &       0.13687218      & 346(4) & 458(5) & 4.2 & 2.4 & 2200 \\  
J303            &               &       $64^3\times192$ &               &       0.137123        &       0.13675466      & 257(3) & 476(5) & 4.2 & 3.2 & 600 \\  
\hline
 \end{tabular} }
\end{centering}
                
        \label{tab:simul}
\end{table}

Our work is based on a subset of the Coordinated Lattice Simulations
(CLS) ensembles with $\Nf = 2+1$ dynamical quarks. They are generated~\cite{Bruno:2014jqa}
using the open-QCD suite\footnote{\tt
  http://luscher.web.cern.ch/luscher/openQCD/}  \cite{Luscher:2012av} and are based on the
O($a$)-improved Wilson-Clover action for fermions, with the parameter
$c_{\rm sw}$ determined non perturbatively in Ref.\ \cite{Bulava:2013cta}, and the tree-level
O($a^2$) improved L\"uscher-Weisz gauge action. The ensembles used in this analysis
were generated at a constant value of the average bare quark
mass such that the improved bare coupling $\tilde g_0$ is kept constant 
 along the chiral trajectory~\cite{Bruno:2014jqa}. In particular, five of the ensembles are 
at the SU(3)-symmetric point, $m_u=m_d=m_s$. The parameters of the
simulations are summarized in Table \ref{tab:simul}.

Results are obtained at four values of the lattice spacing in the
range $a=0.050 - 0.086$\,fm. The scale setting was performed in Ref.\ \cite{Bruno:2016plf}
using a linear combination of the pion and kaon decay constants with a
precision of 1\%. 
The pion masses used in our determination of $\ahvp$ lie in the range $m_\pi\approx 130-420$\,MeV.
% Several ensembles with pion masses below 420\,MeV, including one ensemble at the physical pion mass, are used to perform the chiral extrapolation. 
All the ensembles included in the
final analysis satisfy $m_\pi L > 4$. Furthermore, at two values of the pion mass ($m_\pi
= 280$ and 420\,MeV), two ensembles with the same bare lattice
parameters but different volumes are used to study finite-size
effects. These ensembles with smaller volumes are not included in the
final analysis and are marked by an asterisk in Table \ref{tab:simul}.

All ensembles have periodic boundary conditions (BC) in space.
In the time direction, ensembles E250 and B450 have periodic BCs, 
while all others have open temporal BCs.
% All ensembles have open boundary conditions (BC) in time, except E250 and B450, which have periodic boundary conditions.  
The choice of open boundary conditions was
made in order to address the issue of long auto-correlation times
associated with the topological charge at small lattice
spacing~\cite{Luscher:2011kk}. 
Our use of ensembles with open BCs constitutes part of our motivation 
for employing correlators in the time-momentum representation.
The boundary couples to a tower of states with
vacuum quantum numbers. Therefore, in order to extract vacuum
correlators, sources and sinks of correlation functions should be
placed at a sufficient Euclidean-time separation away from the
boundaries\footnote{In a large volume, the energy of the first excited
state emanating from the boundary is expected to be $2m_\pi$.}. 
On the ensembles with periodic temporal BCs on the other hand, we exploit the translation invariance
in time to increase statistics.

For all ensembles, except E250, the TMR correlation functions
are computed using point sources, randomly distributed in space and in
the center of the lattice in the time direction.  As described in the
next subsection, we use the local vector current at the source and
both the local and the conserved vector currents at the sink.  For the
ensemble E250, propagators are estimated using stochastic sources,
with noise partitioning in spin, colour and
time~\cite{Wilcox:1999ab,Foley:2005ac}. Each source has support on a
single, randomly chosen timeslice. To improve statistics, the TMR
correlator in Eq.\ (\ref{eq:Gx0def}) is averaged over the three spatial
directions. Errors are estimated throughout the calculation using the
jackknife procedure with blocking in order to take into account
auto-correlation effects.

In addition to the direct calculation of the TMR
correlators, the auxiliary calculation of the $\pi\pi$ $I=\ell=1$
scattering phase plays an important role in our determination of
$\ahvp$.  In Ref.\ \cite{Andersen:2018mau}, it has been determined on
ensembles C101, N401, N200, D200 and J303. On all these ensembles except
C101, the pion form factor at timelike kinematics has also been
determined in~\cite{Andersen:2018mau}. 
As compared to the latter reference, 
the number of gauge configurations used for our spectroscopy calculation on 
ensemble D200 has roughly been doubled.
Additionally, we have performed a spectroscopy calculation
on ensemble N203 with a statistics of about 200 gauge configurations.

We have computed the
quark-disconnected contribution to $\ahvp$ on ensembles N401, N203,
N200, D200 and N302. This selection provides us with a handle on the
discretization effects at $m_\pi\simeq 345\,\MeV$ and $m_\pi\simeq 285\,\MeV$, 
and allows us to investigate the chiral behaviour of the disconnected contribution
via the fixed lattice-spacing sequence of ensembles N203, N200, D200.
The disconnected quark loops are computed using four-dimensional,
hierarchically probed noise sources~\cite{Stathopoulos:2013aci} with
512 Hadamard vectors. More technical details on our implementation can
be found in~\cite{Djukanovic:2019jtp}.

\subsection{Lattice correlators, renormalization and O($a$) improvement\la{sec:improvt}}

There are two commonly used discretizations of the vector current in Wilson lattice QCD,
the local and the conserved current. For a single quark flavour $q$, their expressions are 
\ba
V_\mu^{\scriptscriptstyle\rm L}(x) &=& \bar q(x) \gamma_\mu  q(x),
\\
\la{eq:Jdef}
V_\mu^{\scriptscriptstyle\rm C} ( x ) &=& \frac{1}{2} \Big(\bar q ( x + a\hat\mu ) ( 1 + \gamma_\mu ) U_\mu^\dagger ( x )  q ( x ) 
 - \bar q ( x )( 1 - \gamma_\mu ) U_\mu ( x ) q ( x + a\hat\mu ) \Big).
\ea
In our calculation of correlation functions, we always place the local vector current at the origin
in Eq.\ (\ref{eq:Gx0def}); at point $x$, we use either the local or the 
conserved vector current. This provides us with two discretizations of the TMR correlator which share the same continuum limit.
The conserved vector current has the advantage of not undergoing any renormalization or flavour-mixing.

As for the flavour structure, we note that the electromagnetic current can be decomposed in the SU(3) Gell-Mann basis as 
$J_\mu = V_\mu^3 + \frac{1}{\sqrt{3}} V_\mu^8$,
where $V_\mu^a = \bar\psi \gamma_\mu \frac{\lambda^a}{2}\psi$, with $\bar\psi=(\bar u,\;\bar d,\;\bar s)$. Therefore, the 
local current only requires the \emph{non-singlet} renormalization factor $\zv$. The charm-quark contribution is treated separately, at the ``partially
quenched'' level; our treatment of this (small) contribution is described in the next subsection.

We have implemented the Symanzik O($a$) improvement programme as described in Ref.\ \cite{Luscher:1996sc}. 
Since our lattice action is O($a$) improved, we now describe the improvement and renormalization of the vector currents
in order to consistently carry out the Symanzik programme.
The first step is to add to the local vector current an additive O($a$) counterterm with a tuned coefficient $c_{\rm V}^{\,\scriptscriptstyle\rm L}$ 
(respectively $c_{\rm V}^{\,\scriptscriptstyle \rm C}$ for the conserved current) compensating chiral-symmetry violating effects in on-shell correlation functions,
\be\la{eq:Vimp}
(V_\mu^{{\scriptscriptstyle\rm L},a})^{\rm I}(x) = V_\mu^{{\scriptscriptstyle\rm L},a}(x) +  a\,c_{\rm V}^{\,\scriptscriptstyle\rm L}\; \widetilde\partial_\nu \Sigma_{\mu\nu}^a(x),
\ee
where $\widetilde\partial_\nu$ denotes the symmetric lattice derivative\footnote{For the charm, we actually make a different choice described at the end 
of this subsection.} and 
$\Sigma_{\mu\nu}^a(x) = -\frac{1}{2}\bar\psi(x) [\gamma_\mu,\gamma_\nu]\frac{\lambda^a}{2}\psi(x)$.
The second step, which is only required for the local current, is to take into account the following
renormalization pattern,
% apply the mass-dependent renormalization factors,
\ba
\widehat V_\mu^{{\scriptscriptstyle\rm L,3}} = Z_3 (V_\mu^{{\scriptscriptstyle\rm L,3}})^{\rm I},
&\qquad& 
\widehat V_\mu^{{\scriptscriptstyle\rm L,8}} =  Z_8 (V_\mu^{{\scriptscriptstyle\rm L,8}})^{\rm I} + Z_{80} V_\mu^{{\scriptscriptstyle\rm L,0}}.
\ea
We denote by $V_\mu^{{\scriptscriptstyle\rm L,0}} = \frac{1}{2}\bar\psi \gamma_\mu \psi$
 the flavour-singlet current, and the mass-dependent renormalization 
factors are given by~\cite{Bhattacharya:2005rb,Gerardin:2018kpy}
\ba
Z_3 &=& \zv(\tilde g_0)\; (1+ 3\bbarv\; a\mqav + \bv\;a\mql), 
% (g_0,a\mql,a\mqs)
\\ 
Z_8 &=& \zv(\tilde g_0) \Big(1+ 3 \bbarv\; a\mqav + \frac{\bv}{3}\; a(\mql+2\mqs)\Big),
\\
Z_{80} &=& \zv(\tilde g_0)  ({\textstyle\frac{1}{3}}\bv+\fv)\; \frac{2}{\sqrt{3}}a(\mql-\mqs).
\ea
Here $(\mql,\mql,\mqs)$ are the bare subtracted quark masses, $\mqav$ is their average and $\tilde g_0$ is the 
O($a$) improved bare coupling. 
We note that the mixing coefficient $Z_{80}$ is of order $a$ and vanishes in the SU(3)-flavour-symmetric limit.
We use the values of the renormalization factor $\zv$, the critical hopping parameter $\kappa_{\rm crit}$, as well as the improvement coefficients
 $\bv$, $\bbarv$, $c_{\rm V}^{\,{\scriptscriptstyle\rm L}}$ and $c_{\rm V}^{\,{\scriptscriptstyle\rm C}}$,
which are functions of the bare coupling $g_0$, determined recently in~\cite{Gerardin:2018kpy}. 
There it was shown that the obtained values of $\zv$ differ by percent-level 
O($a^2$) effects from an independent high-precision determination \cite{DallaBrida:2018tpn}.
The improvement coefficient $\fv(g_0)$, which is of order $g_0^6$ and only affects the isoscalar contribution 
to $\ahvp$, is neglected. We estimate that the systematic error incurred by this approximation is at present negligible.
Strictly speaking, the connected strange correlator taken in isolation requires an independent, partially quenched improvement coefficient
in the mass-dependent part of the renormalization factor in order to be consistent with O($a$) improvement;
however, we have neglected this effect.

The desired quantity $\ahvp$ is obtained using Eq.\ (\ref{eq:TMRamu}),
where the integral is replaced by a sum over timeslices.  Note that in
the improvement terms entering the TMR correlator, only the temporal
derivative of the tensor current contributes. For the connected light
and strange contributions, we have compared the use of the symmetric
lattice derivative in Eq.\ (\ref{eq:Vimp}) with an alternative implementation
where an integration by parts is used in order to apply the temporal
derivative on the QED kernel $\widetilde K(t)$, and found the
difference to be negligible; therefore we have used the symmetric lattice derivative throughout.
For the charm contribution, however, we have
found it advantageous to use the discrete derivative on the `away'
side, i.e.\ in such a way that the vector-tensor correlator is not
evaluated at a shorter time separation than the vector-vector
correlator itself~\cite{Harris:2015vfa}.  Finally, we remark that we
do not include the O($a^2$) term consisting of the correlation of two
tensor currents.

\subsection{Treatment of the charm contribution}

We treat the charm quark at the partially quenched level: it does not
appear in the simulated action, nor do we include the contribution of
quark-disconnected diagrams containing charm loops. Given that the charm
contribution is about two percent of the total, these approximations
appear fully sufficient at our present level of precision. 

The first task is to tune the value of the bare charm quark mass on
each lattice ensemble.  The mass of the ground state pseudoscalar
$c\bar s$ meson is computed for several values of $\kappa_c$, using
stochastic sources with colour, spin and time dilution. The value of
$\kappa_c$ used in the calculation of $\ahvp$ is then obtained from a
linear interpolation of the squared mass of the lightest $c\bar s$
meson in $1/\kappa_c$ to the point where this mass equals the
experimental value of the $D_s$ meson mass.

We perform a dedicated determination of the multiplicative,
mass-dependent renormalization factor $Z_{\rm V}^c$ for the local
charm current on every lattice ensemble.  The determination is based
on requiring the charm quantum number of the pseudoscalar $c\bar s$ meson to be
exactly unity.  It follows the method used in \cite{Gerardin:2018kpy}
for the light isovector current, and a similar method was already used
in \cite{DellaMorte:2017dyu}. As for the improvement coefficients
$c_{\rm V}^{\,{\scriptscriptstyle\rm L}}$ and $c_{\rm V}^{\,{\scriptscriptstyle\rm C}}$, we use the same values as for
the $u,d,s$ quark flavours. The results for $\kappa_c$ and $Z_{\rm
  V}^c$ are given in Table \ref{tab:resultsSC}, while the individual
pseudoscalar $c\bar s$ meson masses used for the determination of $\kappa_c$ are
collected for reference in Table \ref{tab:charmTableLong} of Appendix~\ref{sec:apda}.

\subsection{Infrared aspects of $\ahvp$: correlator tails, finite-size effects and the chiral limit \la{sec:IR}}

There are a number of aspects of the calculation of $\ahvp$ related to
the long-distance physics of vector correlators that are best
discussed together.  Here, we summarize our understanding of these
issues before applying it to the treatment of lattice data.

In preparation, recall that the TMR correlator can be written, via the
spectral decomposition in finite volume, as the sum of the (positive)
contributions of individual vector states.  In particular, only
isovector vector states contribute in the correlator
\be\la{eq:specsum}
G^{I=1}(t) = \sum_{n=0}^\infty \frac{Z_n^2}{2E_n} e^{-E_n t},
\ee
where the amplitudes $Z_n$ are real, and the discrete, ordered energies $E_n$ are 
real and positive. A similar expression holds for the isoscalar correlator $G^{I=0}(t)$.

\subsubsection*{Controlling the long-time tail of the TMR correlators}

The contribution of the tail of the correlator to $\ahvp$ is enhanced
by the QED kernel. Yet the correlator is affected by a growing
statistical error, as well as a large relative finite-size effect.  We
discuss these two issues in turn.

In order to handle the tail of the correlators, two types of
treatment have been proposed.  Both are based on the fact that at
large Euclidean times, a few terms in the sum of
Eq.~(\ref{eq:specsum}) saturate the correlator to a high degree of
precision, which was one of the motivations for introducing the time-momentum
representation~\cite{Bernecker:2011gh}.  In the first type of
treatment, one explicitly constructs an extension of the correlator for
$t>t_c$, motivated by the spectral representation
(\ref{eq:specsum}). The simplest incarnation of this method, partly
used in our earlier calculation~\cite{DellaMorte:2017dyu}, is to keep
only the lightest of those states and thus to perform a
one-exponential fit to the correlator for Euclidean times around
$t_c$. When a dedicated spectroscopy calculation is available, several
energy levels $E_n$ as well as the overlaps $Z_n$ can be used, so that
the summed contributions of these states already saturate the TMR
correlator at smaller Euclidean times.

A second type of treatment consists in bounding the Euclidean
correlator from above and
below~\cite{CLehnerBounding,Borsanyi:2017zdw,Blum:2018mom}, exploiting
the positivity of the prefactors $Z_n^2/(2E_n)$,
\be\la{eq:bndg}
0\leq G(t_c) e^{-E_{\rm eff}(t_c)(t-t_c)} \leq G(t) \leq G(t_c) e^{-E_N(t-t_c)}, \qquad t\geq t_c,
\ee
where  $N=0$ in the simplest variant,
and $E_{\rm eff}(t)\equiv -\frac{d}{dt}\log G(t)$ is the ``effective mass'' of the correlator.
As a refined variant of this method, a dedicated spectroscopy calculation delivering the energies and matrix elements
of the $N$ lowest-lying states allows one to improve
the control over the tail by applying the bound Eq.\ (\ref{eq:bndg}) to the subtracted correlator
\be\la{eq:Gsub}
\widetilde G(t) = G(t) - \sum_{n=0}^{N-1} \frac{Z_n^2}{2E_n} e^{-E_n t}.
\ee
A challenge one eventually faces in exploiting lattice spectroscopy information is that 
the number of states required to saturate the TMR correlator at a given $t_c$
increases with decreasing pion masses and (roughly proportionally) with the volume.
However, for the ensembles used in this work, the number of states needed is at most four.

\subsubsection*{Finite-size effects on $\ahvp$ in the time-momentum representation}

We now come to the closely related issue of the finite-size effect 
on the observable $\ahvp$ calculated in the time-momentum representation.
At asymptotically large volumes, the finite-size effect is of order $e^{-m_\pi L}$ and can be computed 
in chiral perturbation theory~\cite{Aubin:2006xv,Francis:2013fzp,Aubin:2015rzx}.  At low
pion masses, the leading finite-size effect is expected to come from
the $\pi\pi$ channel, and thus affects the isovector channel only, $G^{I=1}(t)$.
Working in the flavour decomposition of Eq.\ (\ref{eq:Gdecomp}), we take this observation into account 
by applying  $10/9$ of the isovector finite-size correction to the connected light-quark contribution,
and  $-1/9$ of the same correction to the disconnected contribution.

Looking at the finite-size effect on the correlator as a function of
Euclidean time, it has been pointed
out~\cite{Bernecker:2011gh,Francis:2013fzp} that for a given spatial
box size $L$, the tail of the correlator is affected by an
unsuppressed finite-size effect. 
One may define a time $t_i$ beyond which the finite-size effect becomes
sizeable. While $t_i$ grows with $L$, we find that the 
overall finite-size effect on $\ahvp$ is dominated by the tail 
in our present calculation.

\begin{figure}[t!]
        \includegraphics*[width=0.72\linewidth]{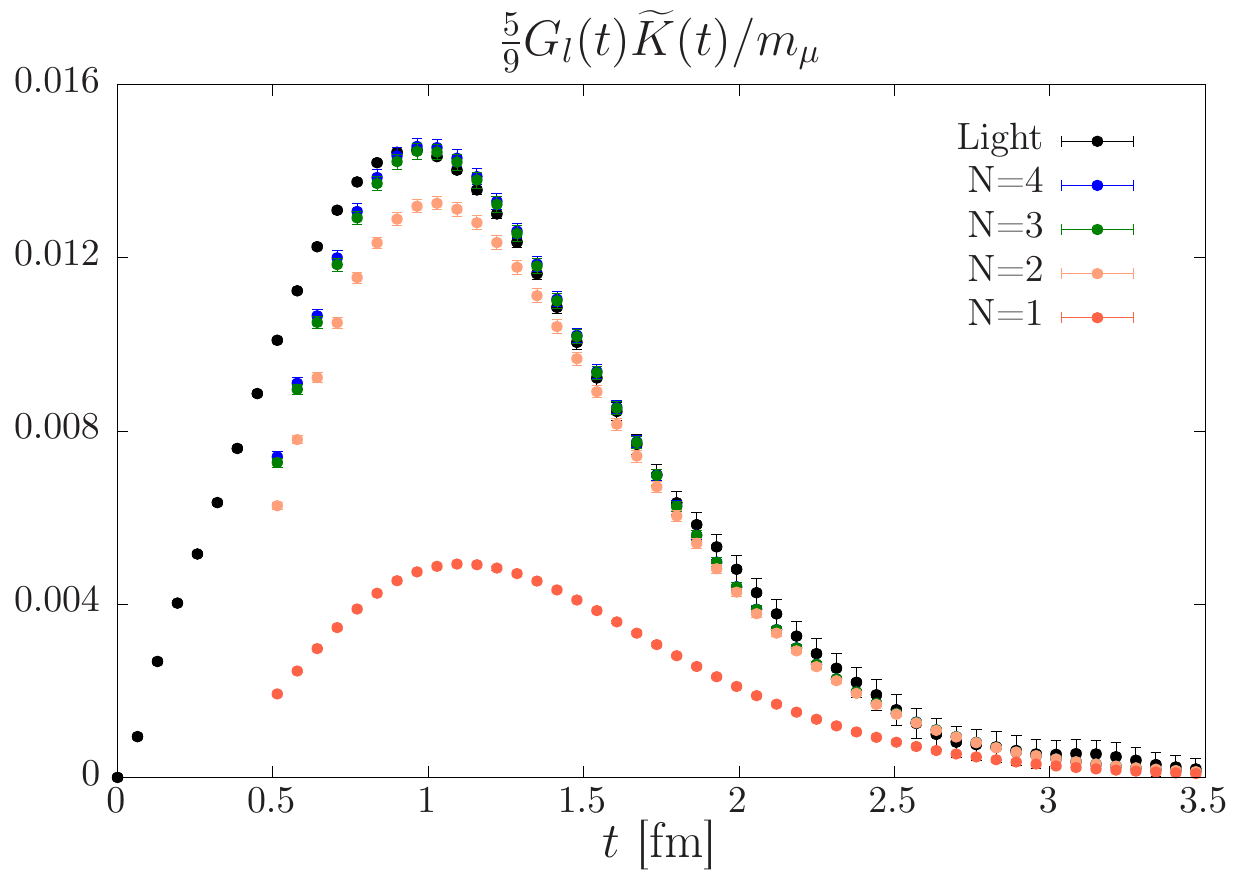}
                
        \caption{Reconstruction of the TMR correlator at long distances using a dedicated spectroscopy analysis on ensemble D200. 
The same gauge configurations are used for the spectroscopy and for the TMR correlator calculation.}
        \label{fig:GEVPD200}
\end{figure}

For $m_\pi L= 4-5$, the tail of the finite-volume isovector correlator
is accurately described by the contribution of a handful of energy
eigenstates; this point will be illustrated in
Fig.\ \ref{fig:GEVPD200}. On the other hand, the tail of the
infinite-volume correlator can be obtained from the timelike pion
factor. Thus, knowledge of this form factor allows one to correct the
tail of the isovector correlator~\cite{Bernecker:2011gh}.  In this
work, we apply the same finite-size correction method as in our
previous calculation~\cite{DellaMorte:2017dyu}, parameterizing the
pion form factor with the Gounaris-Sakurai (GS)
model~\cite{Gounaris:1968mw}.  While too simplistic a model for a
study of the form factor for its own
sake~\cite{Feng:2014gba,Andersen:2018mau}, we expect it to be
sufficient for the purpose of reducing the residual finite-size
effects to a level that is small compared to our current statistical
precision. We emphasize that we only use the GS parametrization of 
the pion form factor for the finite-size
correction, and not for the treatment of the tail of the correlators.

\subsubsection*{The chiral dependence of $\ahvp$}

The TMR correlator for non-interacting pions was given in Ref.\ \cite{Francis:2013fzp}.
For massless pions, it is  given by $G(t) = {1}/(24\pi^2
  |t|^3)$; combined with the asymptotic form of the QED kernel for a
finite muon mass, $\widetilde K(t) \sim 2\pi^2 t^2$, this contribution
generates a logarithmic divergence, which is made finite by a small
but finite pion mass and then yields 
\be\la{eq:amu_chpt1}
\ahvp {\sim} \frac{\alpha^2}{24\pi^2}\log \frac{m_\mu^2}{4m_\pi^2}, 
\qquad m_\pi\to0,~m_\mu {\rm ~fixed}.
\ee
This result and further terms in the expansion have been derived in~\cite{Golterman:2017njs},
where the systematics of the chiral extrapolation has been studied in detail.
The asymptotic form (\ref{eq:amu_chpt1}) only becomes a decent approximation for $m_\pi/m_\mu$ well below 
$1/10$. Thus this logarithmic divergence is largely
irrelevant when describing the pion-mass dependence of $\ahvp$ in the
range $130<m_\pi/{\rm MeV}<300$.
On the other hand, if $m_\mu\ll m_\pi$ and both are small compared to the $\rho$ meson mass,
one finds the leading behaviour
\be\la{eq:amu_chpt2}
\ahvp {\sim} \frac{\alpha^2}{90\pi^2} \frac{m_\mu^2}{4m_\pi^2},
\qquad m_\mu\ll m_\pi\ll m_\rho.
\ee
It turns out that this asymptotic form is rather robust, holding down
to fairly small values of $m_\pi/m_\mu$.  In fact, within the
framework of chiral perturbation theory at next-to-leading
order\footnote{The expression for the momentum-space vector
  correlators at next-to-next-to-leading order can be found
  in~\cite{Golowich:1995kd,Amoros:1999dp}.}  underlying
Eqs.\ (\ref{eq:amu_chpt1}) and (\ref{eq:amu_chpt2}), the combination
$(1+\frac{4m_\pi^2}{m_\mu^2})\ahvp$ only varies by $2\%$ for
 $m_\pi/m_\mu$ in the interval $[1.25,3.0]$ relevant to our lattice
calculations.

At physical quark masses, the overall magnitude of
expression (\ref{eq:amu_chpt2}) is enhanced by the (squared) pion
form factor at timelike kinematics.
In addition, the contribution of the $\pi\pi$ states with
a center-of-mass energy well below the $\rho$-meson mass is numerically
subdominant compared to the resonant contribution.  The $\rho$-meson
mass depends only mildly on the light-quark mass, and thus the steep
behaviour predicted by Eq.\ (\ref{eq:amu_chpt2}) as a function of
$m_\pi$ is superimposed on a larger, more slowly varying contribution.
In our chiral extrapolations, presented in section \ref{sec:phys}, we use these 
observations to construct suitable fit ans\"atze for the chiral extrapolation.

The singular chiral behaviour comes from the isovector channel,
while we expect the isoscalar channel to have a much milder dependence on the pion mass.
Working in the basis of Eq.\ (\ref{eq:Gdecomp}), 
the singular chiral behaviour is split between the connected light-quark contribution 
and the disconnected contribution.
Indeed, in the limit that $m_\mu$ and $m_\pi$ are much smaller than the hadronic scale,
we have $a_\mu^{{\rm hvp,\,disc}}= -\frac{1}{9} a_\mu^{{\rm hvp}}$, and hence, from Eq.\ (\ref{eq:amu_chpt2}),
\be\la{eq:amudiscchiral}
a_\mu^{{\rm hvp,\,disc}} {\sim} -\frac{\alpha^2}{810\pi^2} \frac{m_\mu^2}{4m_\pi^2},
\qquad m_\mu\ll m_\pi\ll m_\rho.
\ee
For orientation, we note that if one inserts the physical pion mass
into this expression, one obtains $a_\mu^{{\rm hvp,\,disc}} =
-10\times 10^{-10}$, and we expect this value to be further enhanced
 by the pion form factor. The important point is that the singular chiral behaviour 
present in the connected light-quark contribution to $\ahvp$ must be present in 
the disconnected contribution as well, with a relative factor of $-1/10$.

\section{Results\label{sec:results}}

In this section we describe the main features of the TMR correlators obtained
on the different lattice ensembles with a view to computing
$\ahvp$. Particular attention is devoted to the correlators at
Euclidean times in the range [1.5, 4.0]\,fm.
In the rescaling of the muon mass, we use the values of $af_\pi$ values
given in Table \ref{tab:resultsL}, corrected for finite-size effects~\cite{Colangelo:2005gd}
and interpolated via a global fit in the pion mass and the lattice spacing.

\subsection{The quark-connected contributions\la{sec:conn}}

\begin{figure}[t!]
        \includegraphics*[width=0.49\linewidth]{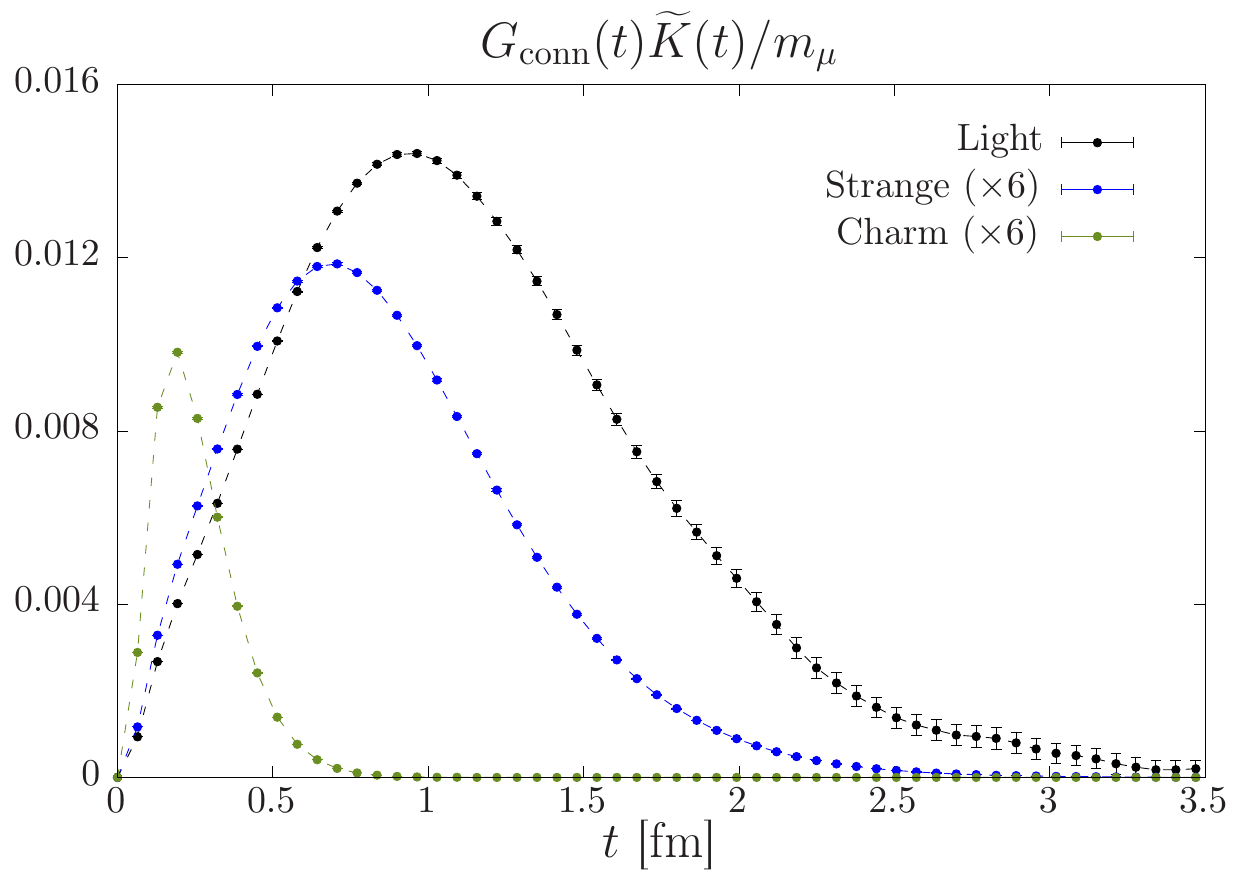}
        \includegraphics*[width=0.49\linewidth]{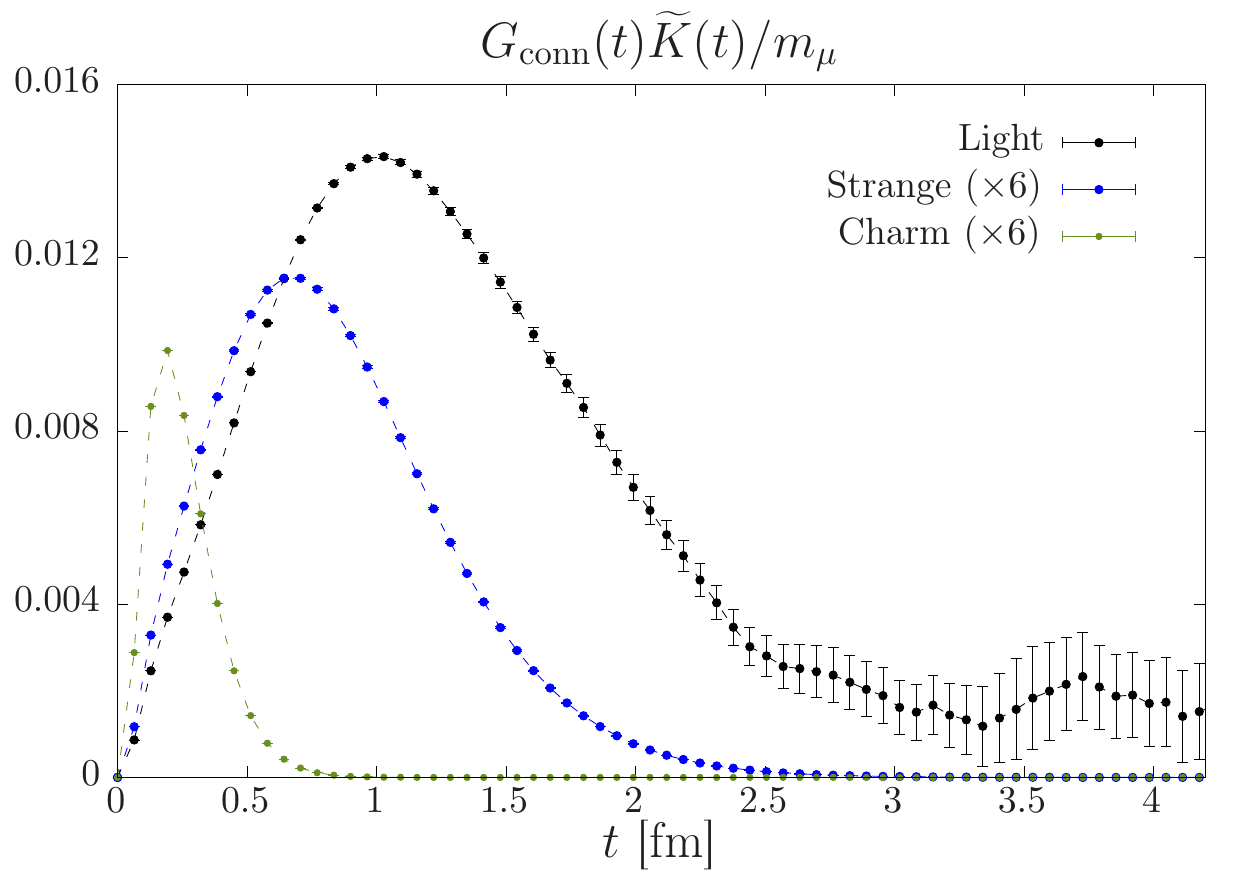}
        \caption{Integrand of Eq.\ (\ref{eq:TMRamu}) in the time-momentum representation for the connected light, 
strange and charm contributions. Left: ensemble D200 with a pion mass of 200 MeV. Right: ensemble E250 at the physical pion mass.
For better visibility, the strange and charm contributions have been scaled by a factor six. The displayed discretization is the local-local one 
 for the light and strange contributions, and the local-conserved one for the charm. 
The muon mass is the $f_\pi$ rescaled one for the light integrand and the physical one for the strange and charm integrands.}
        \label{fig:integrand}
\end{figure}

The integrand of Eq.\ (\ref{eq:TMRamu}) for the connected light,
strange and charm contributions is displayed in
Fig.\ \ref{fig:integrand} for our two ensembles with quark masses
closest to their physical values. The left (right) panel corresponds
to a pion mass of about 200\,MeV (131\,MeV). The light contribution is
clearly very dominant; note that the charm and strange contributions have been scaled by a factor of six 
for better visibility. On a given ensemble, the integrand peaks at
increasingly longer distances as one goes from the charm to the strange to the
light quarks, and the tail becomes more extended. At the same time,
the statistical precision deteriorates. Comparing the left to the
right panel, it is clear that the light contribution becomes harder to
determine with the desired precision as the physical quark masses are approached.
Nevertheless, these plots by themselves do not fully reflect all the known constraints
on the TMR correlator, which is well known to be given by a
sum of decaying exponentials with positive coefficients, as discussed in section \ref{sec:IR}.

Having described the state-of-the-art methods to handle the tail of the correlation function in section \ref{sec:IR},
we now describe how we applied these methods to our data. 
For the strange and charm quark contributions, the TMR correlator is
determined so accurately that practically no particular treatment of
the tail is needed.  We apply the bounding method, Eq.\ \ref{eq:bndg} with $N=0$,
and obtain the results given in Table \ref{tab:resultsSC}.

As for the connected contribution of the light quarks, our
choice for the final analysis is again the bounding method on all
ensembles; the only exception is the physical-pion-mass ensemble E250,
to which we return below.  In applying Eq.\ (\ref{eq:bndg}), we employ
the expression containing the effective mass as a lower bound, and use
as an estimate for the lowest-lying energy level in the channel the
energy obtained by a one-exponential fit to the tail of the TMR
correlator.  On ensemble D200, on which the ground state
lies clearly below the $\rho$ mass and has a relatively weak coupling
to the vector current, we use the auxiliary spectroscopy calculation
to determine its energy. We find it to be close to, but slightly below
the value corresponding to two non-interacting pions, 
$E_0^{\rm free} \equiv 2[\left({2\pi}/{L}\right)^2+m_\pi^2]^{1/2}$.
Table \ref{tab:resultsL} contains our results for the connected contributions of the light quarks.

As discussed in detail in the next subsection,
 the improved statistical precision gained
by exploiting spectroscopic information can be quite significant for light
pion masses, $m_\pi\lesssim200\,$MeV.  Indeed we find that on the
physical mass ensemble E250, on which we do not have direct
spectroscopic information, we cannot achieve a comparable control over
the statistical and systematic error with the simplest variant of the
bounding method. Therefore we proceed as follows.  The isovector
vector energy levels computed on ensembles N203, N200 and D200 allow
us to determine the scattering phase in the $I=\ell=1$ $\pi\pi$
channel~\cite{Andersen:2018mau} for energies up to the four-pion
threshold via the L\"uscher formalism~\cite{Luscher:1991cf}\footnote{See ~\cite{Alexandrou:2017mpi,Fu:2016itp,Guo:2016zos,Wilson:2015dqa,Bali:2015gji} 
for other recent calculations of the scattering phase in the $\rho$ channel.}.
The scattering phase is well described by the effective range formula,
\be\la{eq:effrange}
\frac{k^3}{E}\cot\delta_{11} = \frac{4k_\rho^5}{m_\rho^2 \Gamma_\rho} \Big(1- \frac{k^2}{k_\rho^2}\Big),
\ee
with $k\equiv \frac{1}{2}\sqrt{E^2-4m_\pi^2}$ and $k_\rho$ being the
value of $k$ for $E=m_\rho$.  The parameters $m_\rho$ and
$\Gamma_\rho$ correspond to the $\rho$ meson mass and width.
Furthermore, it has been observed in lattice simulations that
parameterizing the width by 
\be\la{eq:Gamrho}
\Gamma_\rho = \frac{g_{\rho\pi\pi}^2}{6\pi}\,\frac{k_\rho^3}{m_\rho^2},
\ee
 the coupling $g_{\rho\pi\pi}$ only has a weak pion-mass 
dependence.
Therefore, we extrapolate the parameters $(m_\rho,g_{\rho\pi\pi})$
determined on the ensembles N203, N200 and D200 (see Table \ref{tab:spectro}) to obtain their values
for the pion mass corresponding to ensemble E250.
Using these values, we can predict the low-lying energy levels $E_n$ on ensemble E250
by using the L\"uscher correspondence between them and the scattering phase in reverse.
In order to obtain an extension of the TMR correlator on E250, we then fit the 
squared amplitudes $Z_n^2$, given the energy levels. Note that this can be formulated as a linear fit.

In our final choice of parameters, we fit the TMR correlator on E250 in the
interval $26 < t/a < 37$. Then the TMR is summed from
$t=0$ to  $t=28a$ and the multi-exponential extension is used beyond
that time. The numbers given for E250 in Table \ref{tab:resultsL} are
the results from this procedure.

\begin{figure}[t!]
        \includegraphics*[width=0.49\linewidth]{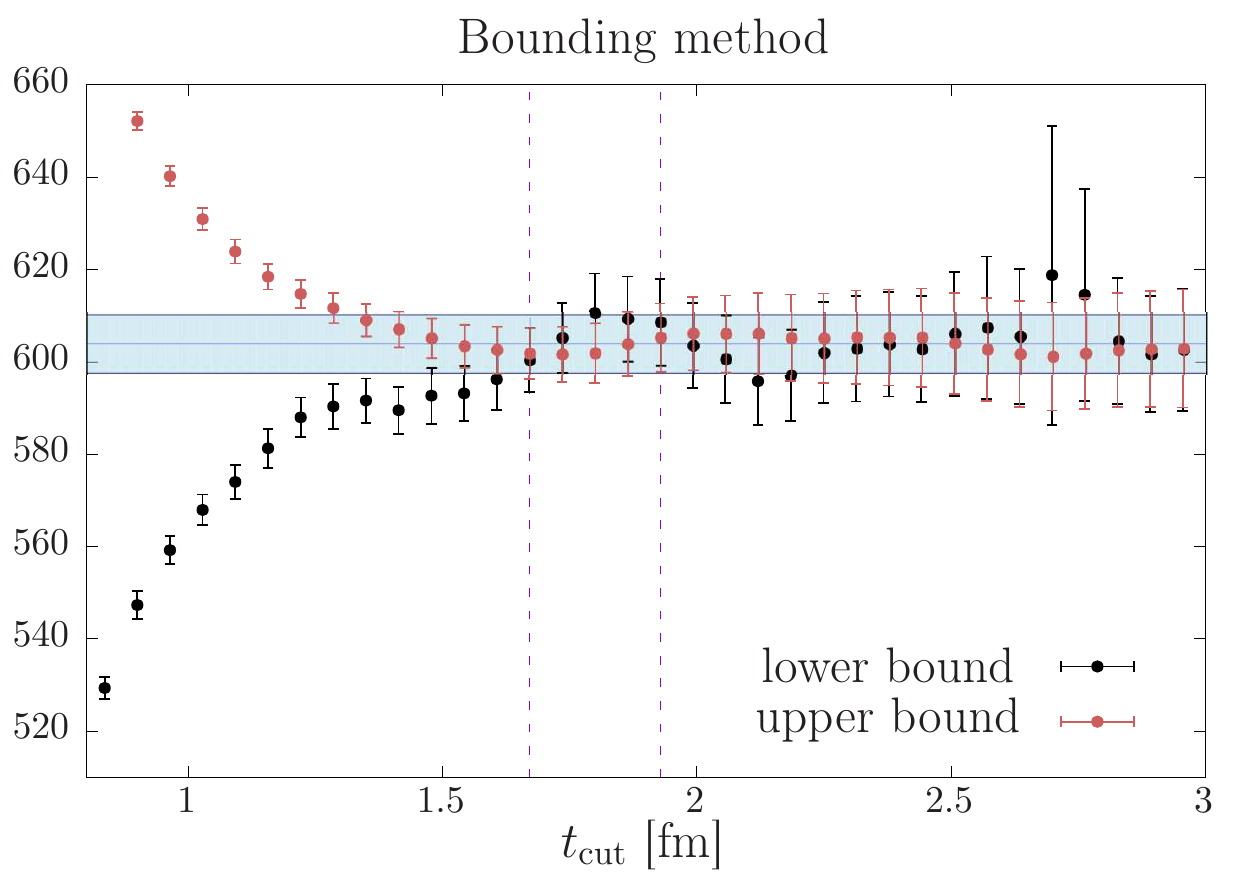}
        \includegraphics*[width=0.49\linewidth]{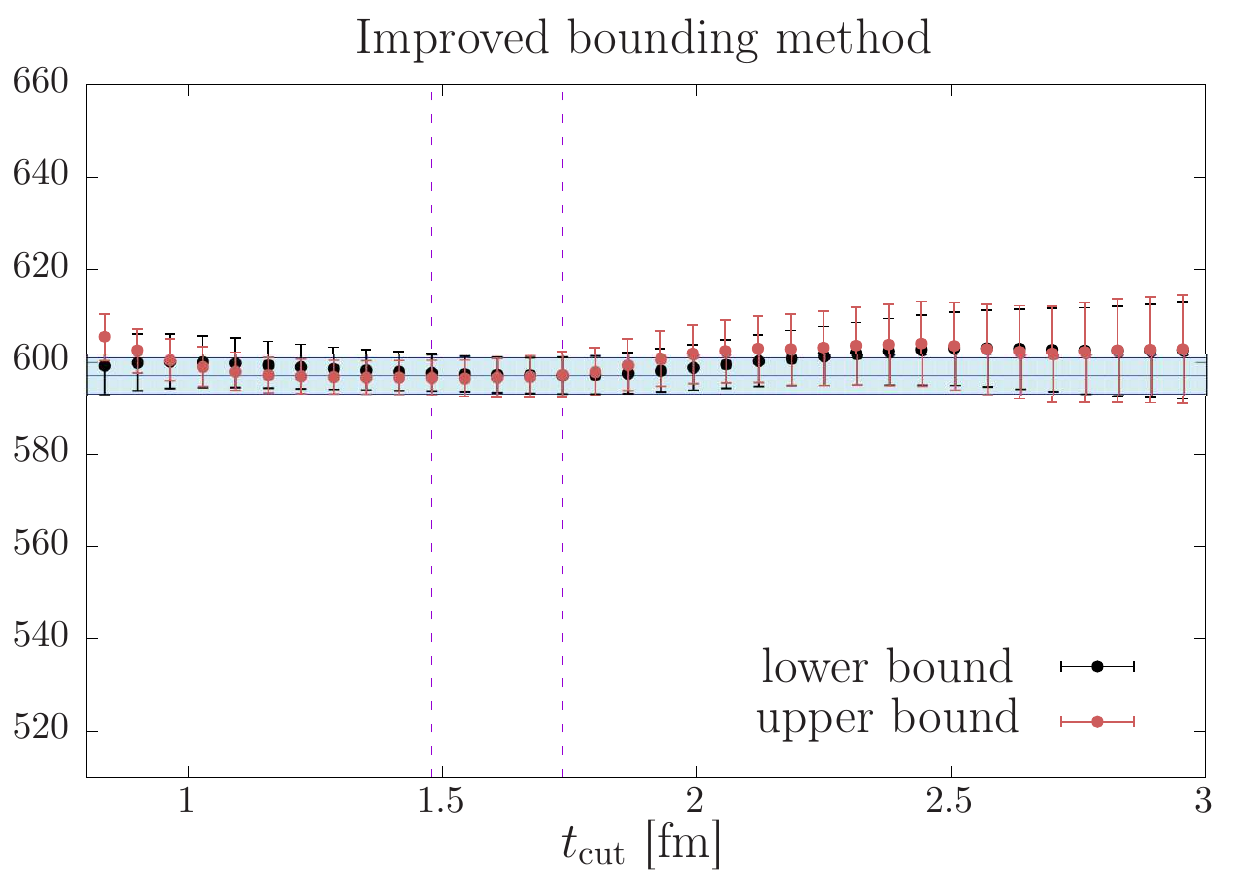}

        \caption{Bounding method with the contribution of $N=0$ (method (1), left) and $N=2$ (method (2), right) states subtracted 
   on ensemble D200 for the local-local correlator and the $f_\pi$-rescaled  muon mass. Results based on 1100 gauge configurations. }
        \label{fig:bounding}
\end{figure}

\subsection{Comparing different methods of extracting $a_\mu^{{\rm hvp},l}$ on ensemble D200 \la{sec:D200}}

\begin{figure}[t!]
\includegraphics*[width=0.49\linewidth]{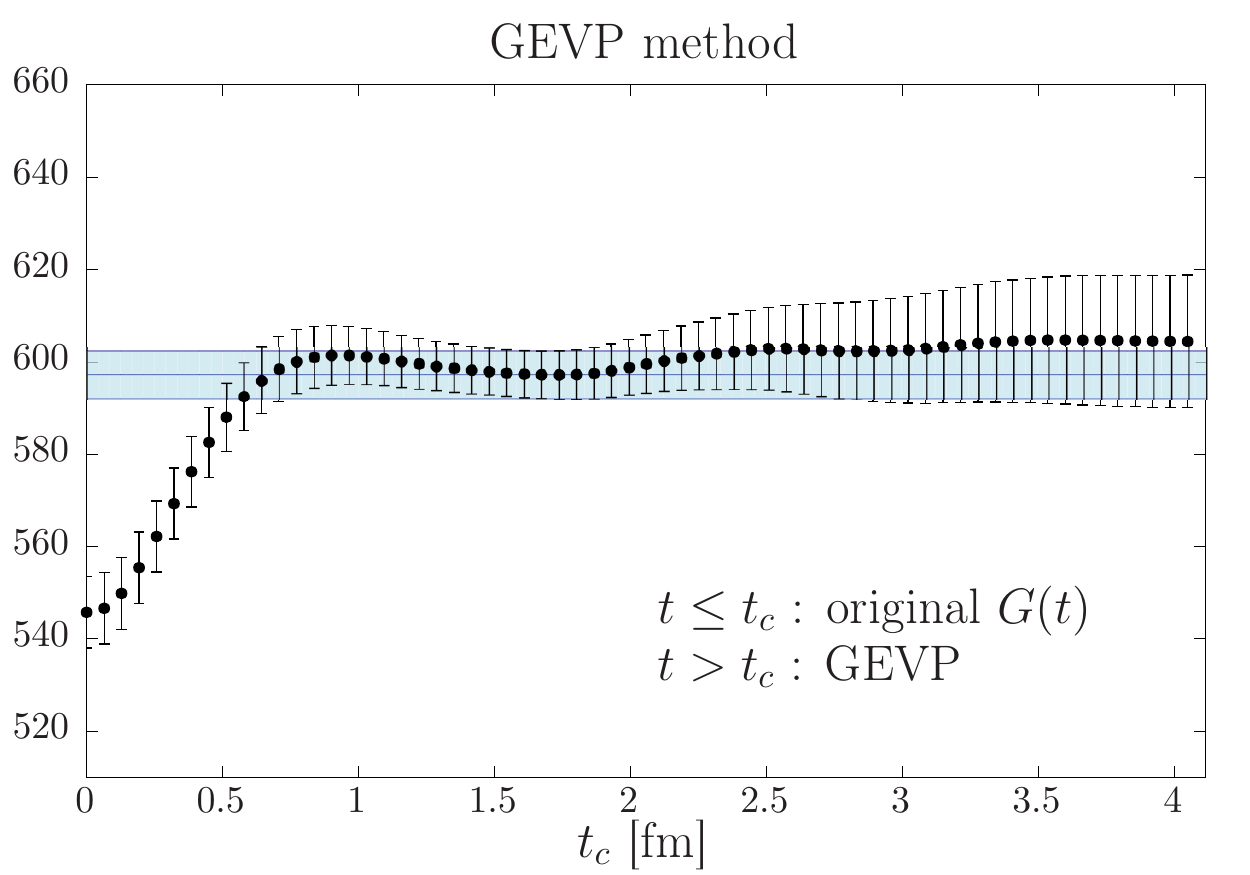}
\includegraphics*[width=0.49\linewidth]{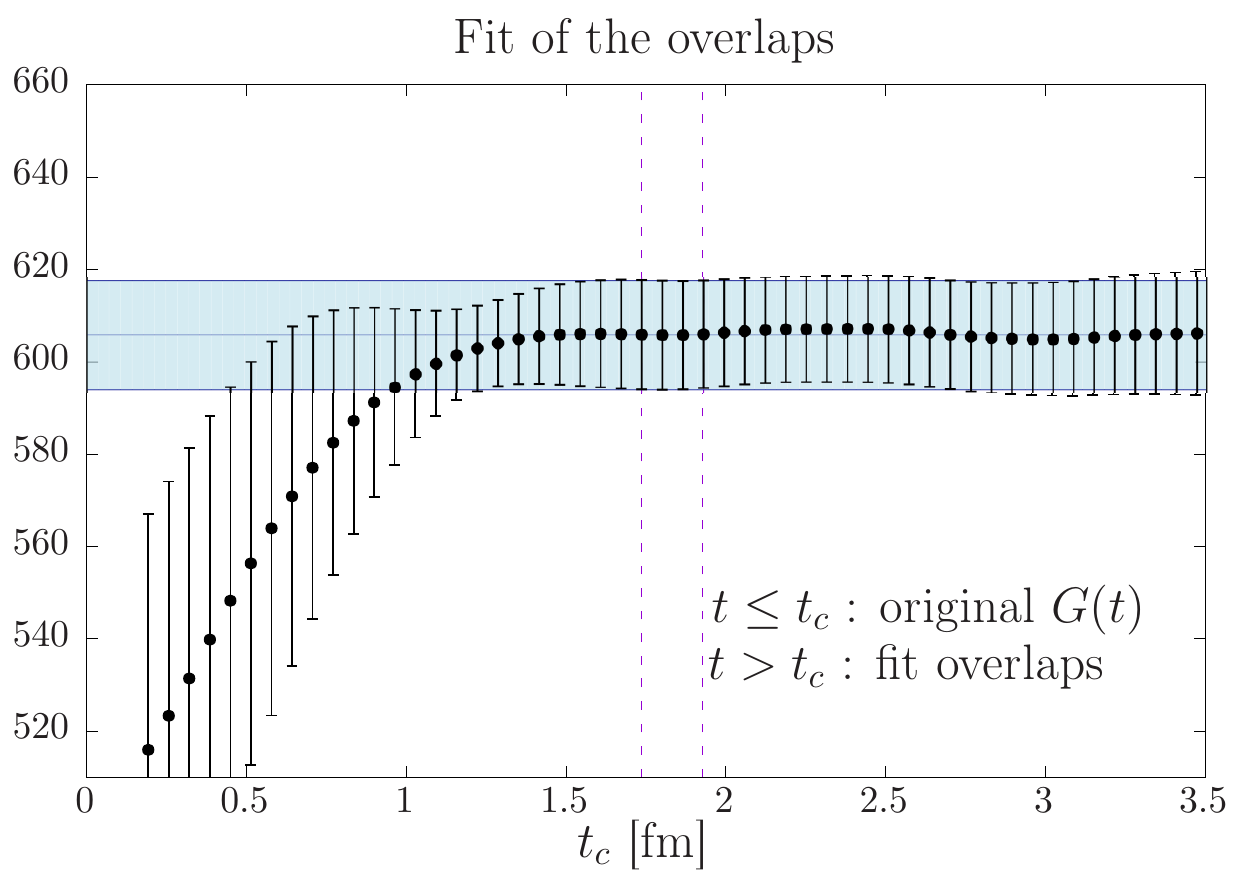}
  \caption{Determination of $a_\mu^{{\rm hvp},l}$ with the $f_\pi$-rescaled muon mass 
  using the extension of the connected light (local-local) correlator
 using $N=2$ energy levels on ensemble D200. 
On the left (method 3), the amplitudes corresponding to energy levels were predetermined in 
a spectroscopy calculation, while on the right (method 4), they are fitted to the TMR correlator.
Results based on 1100 gauge configurations.}
        \label{fig:D200tail}
\end{figure}

On ensemble D200 at $m_\pi=200\,$MeV, we have detailed information on
the scattering phase and the timelike pion form factor.  We can thus
test the validity of the procedure we applied on the physical
pion-mass ensemble E250, described in the previous subsection.

Thus on D200 we apply and compare four different methods to handle the tail of the 
light connected correlator:
\begin{enumerate} 
\item[(1)] the bounding method without subtractions ($N=0$);
\item[(2)] the bounding method after subtracting the contribution of $N=2$ states;
\item[(3)] the extension of the correlator using the auxiliary information on the first two energy levels $E_n$
and their amplitudes $Z_n$;
\item[(4)] the extension of the correlator using the auxiliary information on the first two energy levels $E_n$,
but fitting the amplitudes to the TMR correlator.
\end{enumerate}
One motivation for comparing these particular methods is that on E250, we cannot apply the second or third method,
while the first method would result in a large statistical error. Therefore, we apply the last method on E250, and 
presently test whether it gives consistent results on ensemble D200.

Fig.\ \ref{fig:bounding} compares the results for $\ahvp$ from methods
(1) and (2), as a function of the time $t_c$ at which the upper and
lower bounds start to be used instead of the TMR correlator itself.
The values are consistent with each other, however method (2) yields a
significantly reduced statistical uncertainty. This outcome is not
surprising, since important auxiliary information is used in method
(2).

A comparison of methods (3) and (4) is shown in
Fig.\ \ref{fig:D200tail}, showing the resulting $\ahvp$ as a
function of the time $t_{c}$ at which the TMR correlator is
replaced by the multi-exponential extension. The result of method (4)
is consistent with that of method (3), albeit with an enlarged
statistical uncertainty.  In addition we have checked that the values
of the amplitudes of the first two states as extracted from the fit in
method (4) are well consistent with their direct spectroscopic
determination. Table \ref{fig:D200methods} presents the results
obtained on D200 with the four different methods.

\begin{table}
\caption{Dependence of the D200 result for 
$10^{10}\times a_\mu^{{\rm hvp},l}$ 
on the methods described in the text, using the local-local TMR correlator.}
\vskip 0.1in

\begin{tabular}{c@{~~~}c@{~~~}c}
\hline
Method &  No rescaling  &  With $f_\pi$ rescaling \\
\hline
1.      &  605.9(6.3)  & 604.2(7.4)  \\
2.      &  599.0(4.0)  & 597.1(5.2)   \\
3.      &  599.4(3.9) &   597.7(5.0)     \\
4.      & 607.7(11.8) &   605.7(11.8)     \\
\hline
\end{tabular}
\la{fig:D200methods}
\end{table}

\subsection{Finite-volume effects\label{sec:FSE}}

\begin{figure}[t!]
\includegraphics*[width=0.72\linewidth]{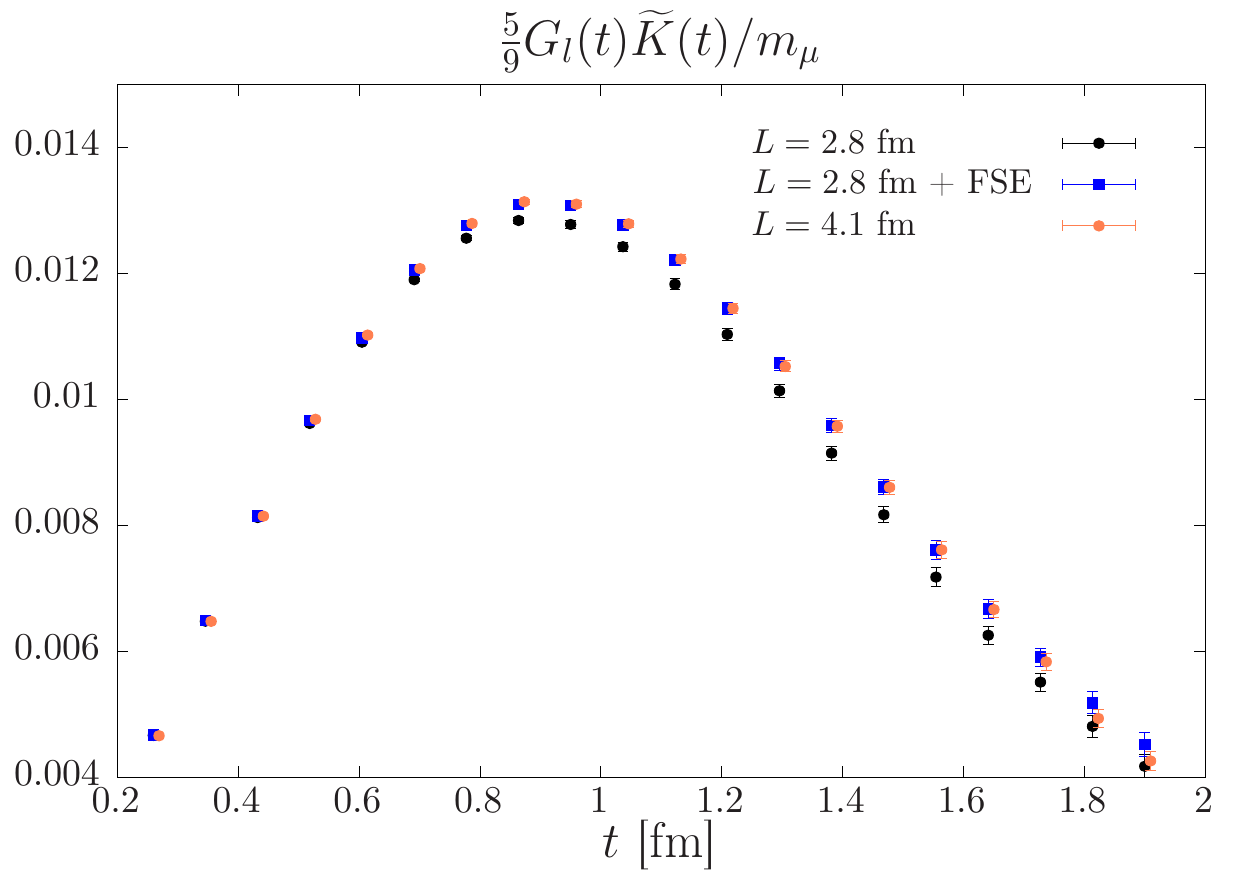}
      \caption{Testing the finite-size correcting procedure described in the main text
               on the ensembles N101 and H105 at a pion mass of 280\,MeV. 
               The scale-setting uncertainty is not displayed, since both ensembles have the same lattice spacing.
}
        \label{fig:FSE}
\end{figure}

As explained in section \ref{sec:IR}, in the isospin basis, we would
correct the $I=1$ correlator for finite-size effects stemming from the
$\pi\pi$ states, and neglect such effects on the $I=0$ correlator. However,
we work in the basis of Eq.\ (\ref{eq:Gdecomp}). In this basis, such a correction
corresponds to applying an additive finite-size correction to the
connected light contribution ($\frac{5}{9}G_l(t)$), weighted by a
factor of $10/9$ relative to the correction of the $I=1$
correlator. At the same time, the disconnected contribution $G_{\rm
  disc}$ must be corrected by $-1/9$ of the $I=1$ correction.  It is
indeed well known that the tail of $G_{\rm disc}(t)$ is given by
$(-1/9)G^{I=1}(t)$~\cite{Francis:2013fzp}.

The $I=1$ finite-size corrections are given in Table \ref{tab:FSE} for
every ensemble. They are computed as in~\cite{DellaMorte:2017dyu},
assuming a GS parametrization of the pion form factor.  However, in
contrast to~\cite{DellaMorte:2017dyu}, the parameters of the GS
parametrization are obtained either by fitting the tail of the TMR
correlator using the relations between the $(E_n,Z_n)$ and the pion
form factor~\cite{Luscher:1991cf,Meyer:2011um}, or by using the
results for $m_\rho$ and $g_{\rho\pi\pi}$ from a dedicated pion form
factor calculation, when available.  This concerns ensembles C101,
N401, N203, N200, D200 and J303.

We have neglected finite-size effects for the connected strange
contribution, except for the SU(3) symmetric ensembles, where
finite-size effects are the same as for the light-connected
contribution\footnote{At the SU(3) symmetric point, the isovector
  correlator receives an additional finite-size correction due to kaon
  loops, which amounts to half the correction due to the pion loop.}.
Similarly, no finite-volume correction is applied to the charm-quark
contribution.

We have performed a direct lattice calculation of the FSE on two ensembles,
N101 and H105, with different volumes, $L=2.8\,$fm and 4.1\,fm, at a
common pion mass of 280\,MeV.  Figure~\ref{fig:FSE} shows that a
finite-size effect is clearly visible and statistically significant.
After the finite-size correction obtained via the GS model for the
pion form factor, the two correlators are in excellent agreement. This
test gives us confidence that the finite-size correction we apply is
reliable at our level of statistical precision.

\begin{figure}[t!]
\includegraphics*[width=0.64\linewidth]{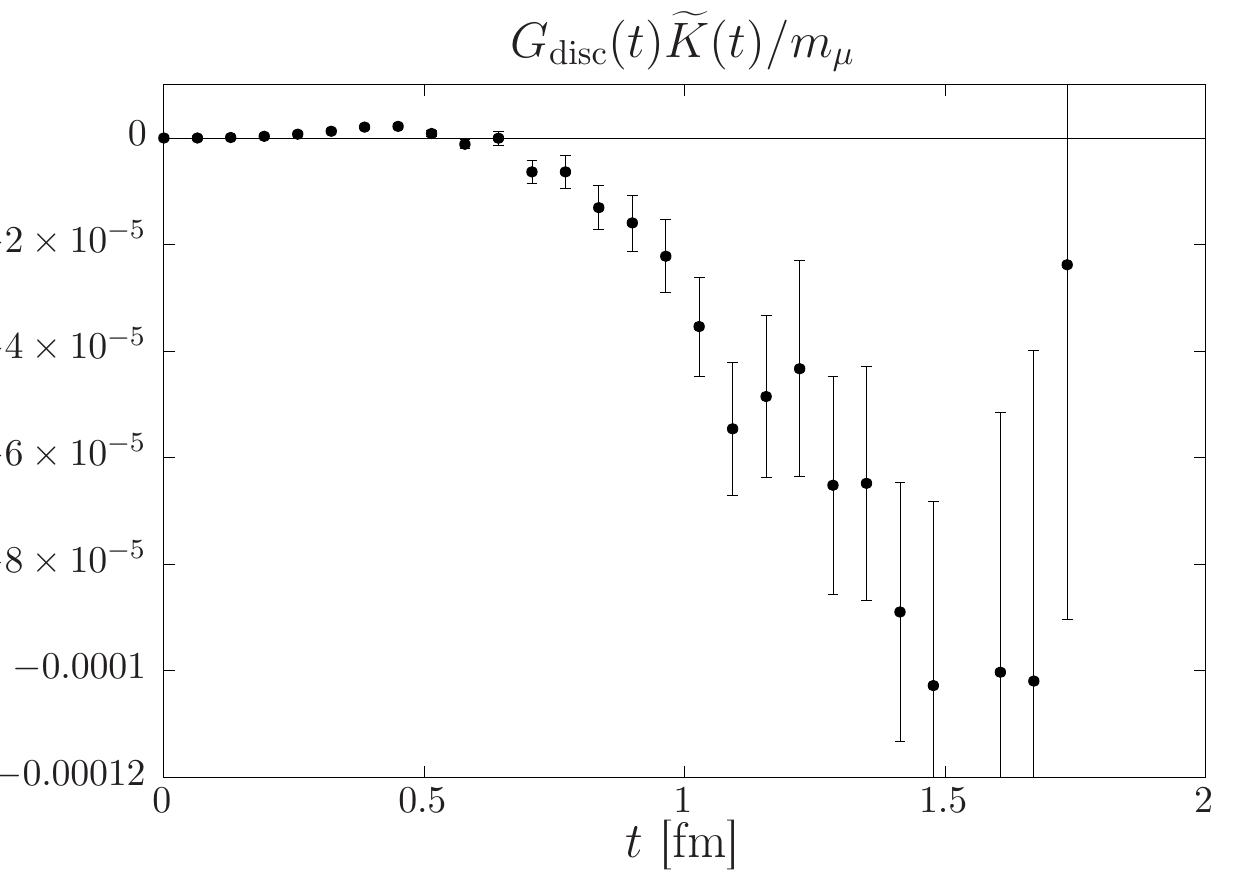}
      \caption{Integrand of Eq.\ (\ref{eq:TMRamu}) in the time-momentum representation for the disconnected contribution on ensemble N200, using the local-local discretization and the physical muon mass.
}
        \label{fig:disc_intgd}
\end{figure}

\subsection{The quark-disconnected contribution}

We have computed the quark-disconnected contribution on a number of
lattice ensembles, namely H105, N401, N203, N200, D200, N302.  A
typical integrand is shown in Fig.\ \ref{fig:disc_intgd}.  The signal
for the quark disconnected contribution is lost around $t=1.5\,$fm.
Given that the \emph{absolute} error of the integrand for $\ahvp$
grows asymptotically, it is clear that additional information
constraining the tail of the disconnected TMR correlator is mandatory.

We have therefore adopted the following strategy. In our $\Nf=2+1$ simulations, 
the isoscalar correlator $G^{I=0,c\!\!/}(t)$ of the $(u,d,s)$ quarks\footnote{The notation $G^{I=0,c\!\!/}$ is introduced
to distinguish this correlator from the full isoscalar contribution $G^{I=0}$, which also contains the charm contribution.} 
admits a positive spectral representation
analogous to Eq.\ (\ref{eq:specsum}), with positive prefactors
multiplying the exponentials. We expect that on the ensembles on which we have
computed the disconnected diagrams, the dominant exponential in a large window of Euclidean times
corresponds to the $\omega$ meson mass.
As we did not perform a dedicated calculation of the $\omega$ mass, 
we use our determination of the $\rho$ resonance mass. 
Since the latter is slightly lower than the $\omega$ mass, this  is a conservative choice.
We can therefore apply the bounding method in the following form,
\be
0\leq G^{I=0,c\!\!/}(t) \leq G^{I=0,c\!\!/}(t_c) e^{-m_\rho(t-t_c)}, \qquad t\geq t_c.
\ee
In order to quote a value $a_\mu^{\rm hvp,disc}$ for the quark-disconnected contribution to $\ahvp$,
we subtract the connected light and strange contributions from the isoscalar contribution $a_\mu^{{\rm hvp},I=0,c\!\!/}$,
\be
a_\mu^{\rm hvp,disc} =  a_\mu^{{\rm hvp},I=0,c\!\!/} - \frac{1}{10}  a_\mu^{{\rm hvp},l} -  a_\mu^{{\rm hvp},s}.
\ee
Our results for $a_\mu^{\rm hvp,disc}$ are listed in Table \ref{tab:disc} in Appendix \ref{sec:ResTabs}.

\section{Results at the physical point\label{sec:phys}}

Having determined the various contributions to $\ahvp$ on a number of 
gauge ensembles, we proceed to extrapolate these results to the continuum
and to the physical pion mass, $m_\pi=134.97\,$MeV.
We use as chiral expansion variable the dimensionless ratio
\be
\widetilde y = \frac{m_\pi^2}{16\pi^2 f_\pi^2},
\ee
where $m_\pi$ and $f_\pi$ have been determined on each ensemble.

\subsection{The connected strange and charm contributions}

\begin{figure}[t!]
    \includegraphics*[width=0.49\linewidth]{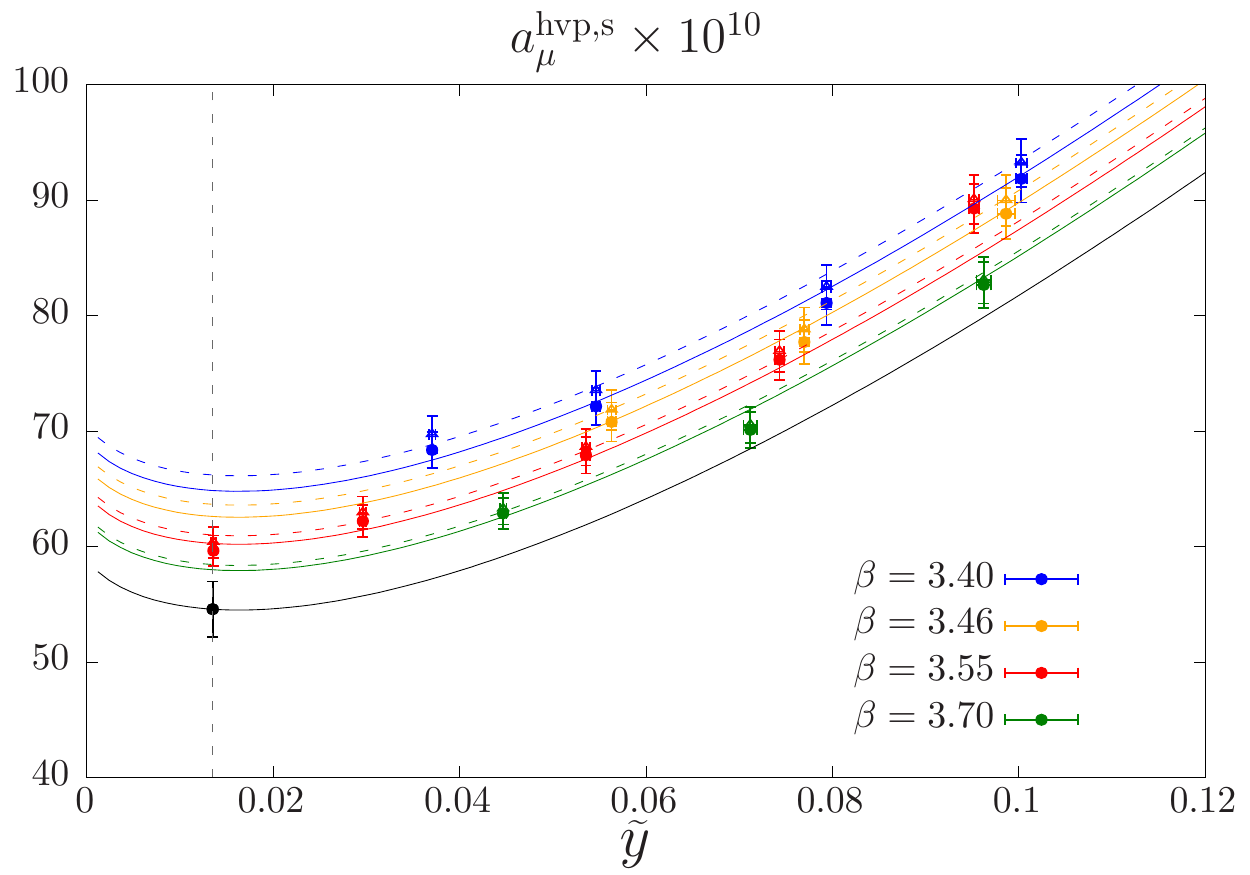}
    \includegraphics*[width=0.49\linewidth]{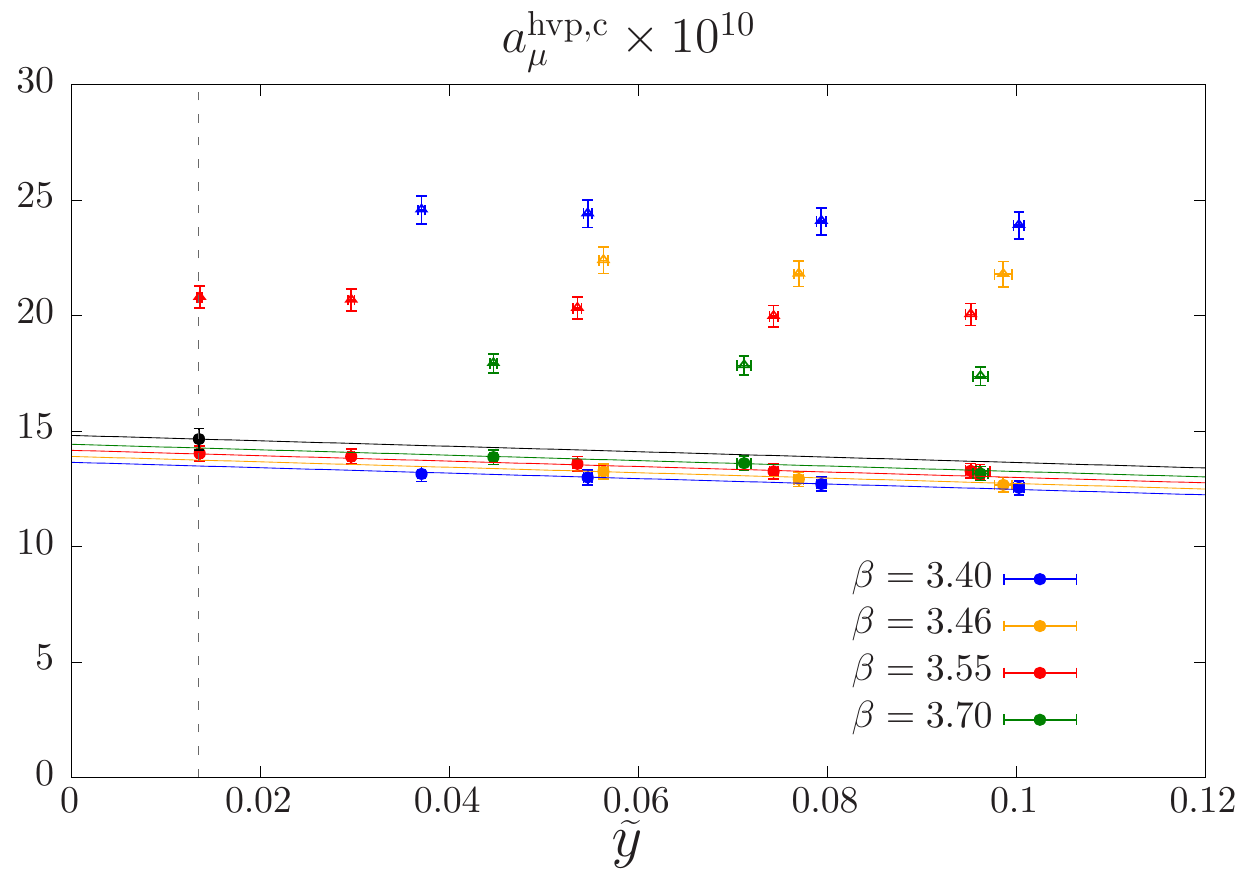}

        \caption{Extrapolation of the connected strange and charm contributions to $\ahvp$ with a muon mass fixed to its physical value.
The black curve represents the chiral dependence in the continuum, and the black point the final result at the physical pion mass.}
        \label{fig:amusc}
\end{figure}

For the strange-quark contribution, the statistical error 
(excluding the lattice spacing uncertainty) is below 1\% for all the
ensembles, and in many cases below 0.5\%, typically for those ensembles with close-to-physical quark masses.
See Table \ref{tab:resultsSC}.
The error is therefore dominated by the scale-setting uncertainty, which enters through the combination $tm_\mu$
in the integrand (\ref{eq:TMRamu}).
We extrapolate the results of the individual ensembles to the physical point using the fit ansatz 
\be 
a_\mu^{{\rm hvp},s}(a,\widetilde y,d) = a_\mu^{{\rm hvp},s}(0,\widetilde y_{\rm exp}) 
 + \delta_d\;a^2 + \gamma_1 (\widetilde y - \widetilde y_{\rm exp} ) 
+ \gamma_2 \,(\widetilde y\log{\widetilde y}- \widetilde y_{\rm exp} \log \widetilde y_{\rm exp}).
\ee 
The index $d$ labels the discretization, local-local or local-conserved.
We observe a rather mild continuum extrapolation and both discretizations are in very good
agreement. The fit goes perfectly through our physical mass ensemble
and our final result for the connected strange-quark contribution is
\be
a_\mu^{{\rm hvp},s} = (54.5\pm 2.4\pm 0.6)\times 10^{-10},
\ee
where the first error is statistical and the second is the
systematic error from the chiral extrapolation. The latter
 is estimated from the difference between the results
obtained if one includes or excludes ensembles with $m_\pi>300\,$MeV.
The chiral and continuum extrapolation is illustrated in the left panel of Fig.\ \ref{fig:amusc}.

For the charm-quark contribution, the statistical error is below 0.3\%
for all the ensembles, and the error on the tuning of the charm
hopping parameter is of similar magnitude.  The error is again
dominated by the scale-setting uncertainty.  As can be seen on the
right panel of Fig.\ \ref{fig:amusc}, the lattice discretization of
the correlator using two local vector currents leads to large cut-off
effects: we observe a discretization effect of almost 70\% at our
coarsest lattice spacing. By contrast, for the local-conserved discretization the
discretization effect is only 8\%. Thus we prefer not to use the
local-local discretization in our continuum
extrapolation of the connected charm contribution. Furthermore, the data also suggest a very flat chiral
behaviour, and we therefore use the fit ansatz
\be
a_\mu^{{\rm hvp},c}(a,\widetilde y) = a_\mu^{{\rm hvp},c}(0,\widetilde y_{\rm exp}) + \delta\;a^2 + \gamma_1 (\widetilde y - \widetilde y_{\rm exp} ).
\ee
At the physical point, we obtain
\be
a_\mu^{{\rm hvp},c} = (14.66\pm 0.45 \pm 0.06)\times 10^{-10},
\ee
where the first error is statistical and the second is the systematic error induced by the chiral extrapolation.
The chiral and continuum extrapolation is illustrated in Fig.\ \ref{fig:amusc} (right panel).

A comparison of the strange and charm contributions to $\ahvp$ with recent publications is shown in Fig.\ \ref{fig:amucomp}.

\subsection{The connected light-quark contribution}

\begin{figure}[t!]
        \includegraphics*[width=0.49\linewidth]{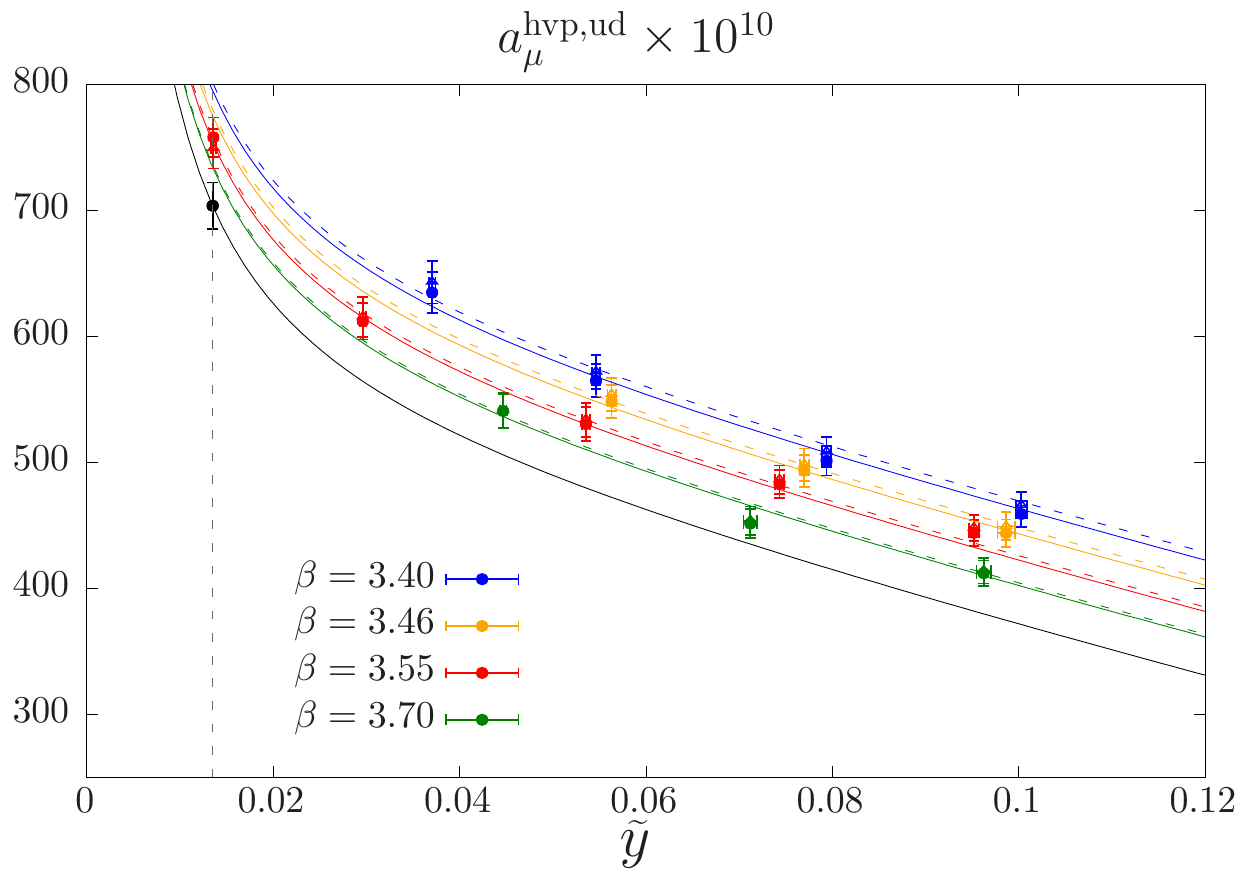}
\includegraphics*[width=0.49\linewidth]{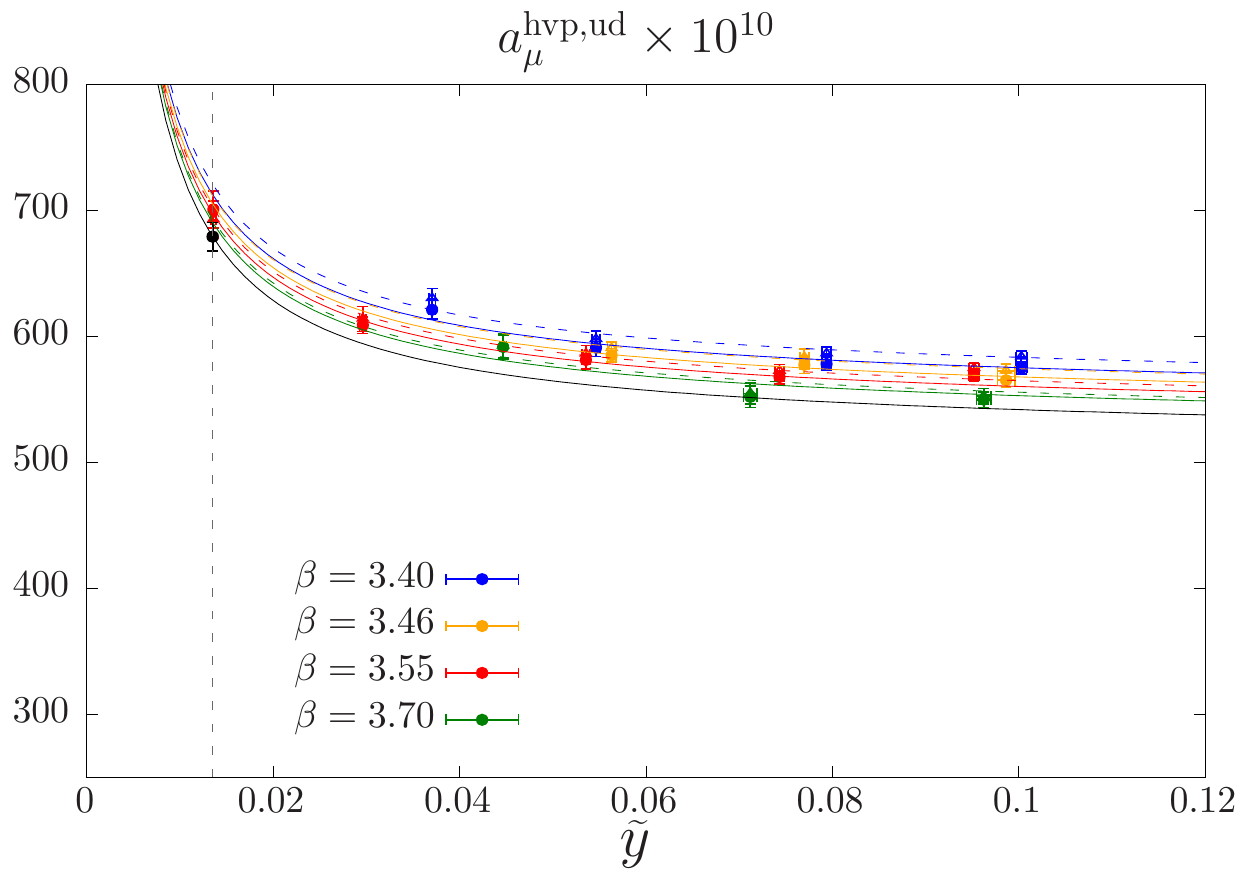}

        \caption{Extrapolation of the connected light contribution to $\ahvp$, using 
 the physical value of the muon mass in the kernel $\widetilde K(t)$ on all ensembles (left panel), 
 and using the rescaled mass $m_\mu^{\rm phys}\cdot \frac{f_\pi^{\rm latt}}{f_\pi^{\rm phys}}$ (right panel).
 The result of the fit based on Eq.\ (\ref{eq:extrap_1ovy}) is shown. The black curve represents the chiral 
dependence in the continuum, and the black point the final result at the physical pion mass.
 }
        \label{fig:amul}
\end{figure}

We have achieved a statistical error of just over two percent on
$a_\mu^{{\rm hvp},l}$ on the physical-mass ensemble E250, and of
$1.0-1.2\%$ on all other ensembles. An important role of the other
ensembles is to constrain the continuum limit, which would be very
costly to achieve directly at the physical pion mass.  Our lattice
data points are displayed as a function of $\widetilde y$ in
Fig.~\ref{fig:amul}, with and without the rescaling of the muon mass
with $f_\pi$.  We observe that the rescaled data on the right panel
has a reduced dependence on $\widetilde y$, as well as on the lattice
spacing. We therefore decide to use the rescaled data for our primary analysis,
but also perform the analysis of the unrescaled data in parallel for comparison.

The expected chiral behaviour of the light connected contribution is reviewed in section \ref{sec:IR}.
Taking into account these considerations, we have used the following ans\"atze to simultaneously extrapolate our results to the continuum
and to physical quark masses:
\begin{subequations}
\begin{align}
a_\mu^{{\rm hvp},l}(a,\widetilde{y},d) &=a_\mu^{{\rm hvp},l}(0,\widetilde{y}_{\exp}) + \delta_d \, a^2 + \gamma_1 \, \left( \widetilde{y} -  \widetilde{y}_{\exp} \right)  + \gamma_2 \, \left( \log \widetilde{y} -  \log \widetilde{y}_{\exp} \right), \label{eq:extrap_log} \\
a_\mu^{{\rm hvp},l}(a,\widetilde{y},d) &=a_\mu^{{\rm hvp},l}(0,\widetilde{y}_{\exp}) + \delta_d \, a^2 + \gamma_3 \, \left( \widetilde{y} -  \widetilde{y}_{\exp} \right)  + \gamma_4 \, \left( \widetilde{y}^2 -  \widetilde{y}_{\exp}^2 \right), \label{eq:extrap_ys} \\
a_\mu^{{\rm hvp},l}(a,\widetilde{y},d) &=a_\mu^{{\rm hvp},l}(0,\widetilde{y}_{\exp}) + \delta_d \, a^2 + \gamma_5 \, \left( \widetilde{y} -  \widetilde{y}_{\exp} \right)  + \gamma_6 \, \left( 1/\widetilde{y} - 1/\widetilde{y}_{\exp} \right), \label{eq:extrap_1ovy} \\
a_\mu^{{\rm hvp},l}(a,\widetilde{y},d) &=a_\mu^{{\rm hvp},l}(0,\widetilde{y}_{\exp}) + \delta_d \, a^2 + \gamma_7 \, \left( \widetilde{y} -  \widetilde{y}_{\exp} \right)  + \gamma_8 \, \left(\widetilde{y} \log \widetilde{y} -  \widetilde{y}_{\exp} \log \widetilde{y}_{\exp} \right), \label{eq:extrap_ylog}
\end{align}
\end{subequations}
where  $d$ is a label for the local-local or local-conserved correlator.
All ans\"atze contain four parameters to be fitted, including an O($a^2$) term to account for discretization errors.
Ansatz (b) assumes a purely polynomial behaviour in the variable $\widetilde y$, 
while fit (d) allows for a non-analytic $\widetilde y \log\widetilde y$ term. The latter ansatz was used in our previous $N_{\rm f}=2$ 
calculation~\cite{DellaMorte:2017dyu}. Ans\"atze (a) and (c) are directly motivated by the discussion in section \ref{sec:IR},
(a) containing the logarithmic singularity that appears in the limit $m_\pi\to 0$ at fixed muon mass,
while (c) contains the  $1/m_\pi^2$ term relevant in the regime $m_\mu\ll m_\pi\ll m_\rho$.

We give the results we obtain from these four ans\"atze, with and without rescaling $m_\mu$, in Table  \ref{tab:fit}.
We have performed these fits either including all ensembles, or imposing cuts on $\widetilde y$, corresponding to pion masses
below 360\,MeV or, alternatively, below 300\,MeV. 
Focusing first on the rescaled data, we note that fits (a), (c) and (d) yield  $\chi^2/{\rm d.o.f.}\approx 1.0$
while fit (b) produces higher values of around 1.6.
With the pion-mass cut at 360\,MeV,  one sees that results (a) and (c) show good consistency 
and yield somewhat larger values of $a_\mu^{{\rm hvp},l}$ than fits (b) and (d).
Given the more singular chiral behaviour of  ans\"atze (a) and (c), this outcome is not unexpected.
Looking at the stability of the final value for $a_\mu^{{\rm hvp},l}$ as a function of the pion-mass cut, 
we observe excellent stability in the case of fits (a) and (c), while the results of fits (b) and (d) systematically
drift upward as a stronger pion-mass cut is imposed. With the strongest cut, $m_\pi<300\,$MeV, all four ans\"atze
yield the same result within half a standard deviation. In view of the greater stability of fits (a) and (c) against
pion-mass cuts, and the stronger theoretical motivation underlying them, 
we choose to average the results of fit (a) and (c) with the cut $m_\pi<360\,$MeV for our final central value.
As a systematic error, we take the full difference between the results of these fits, and thus our final 
result for the connected light-quark contribution is
\be \la{eq:amulight_final}
a_\mu^{{\rm hvp},l} = (674\pm 12 \pm 5)\times 10^{-10}.
\ee

A few further remarks are in order. 
It is important to note that the results of fits (a) and (c) are in very good agreement with the values of $a_\mu^{{\rm hvp},l}$ 
directly obtained on ensemble E250 with the rescaled muon mass; see Table \ref{tab:resultsL}. 
We also remark that the statistical uncertainty on the final result Eq.~(\ref{eq:amulight_final})
is only 20\% lower than the statistical uncertainties on E250; we conclude that the chiral extrapolation of our results obtained at heavier
pion masses, which tend to be more precise, does not lead to an artificially small final uncertainty.
A comparison with the extrapolated results obtained from the standard kernel, shown in the left part of Table~\ref{tab:fit},
shows that the latter lie systematically higher than the rescaled ones. Their statistical uncertainty is larger by about 50\%
 than in the unrescaled case.
 Still, when combining statistical and systematic uncertainties in quadrature of Eq.\ (\ref{eq:amulight_final}),
the central value of fit (c) only lies 1.6 standard deviations higher than our final central value Eq.\ (\ref{eq:amulight_final}).

\begin{table}[t]
\caption{Results of the connected light-quark contribution in units of $10^{-10}$ using different fits and cuts. Left: using the standard kernel. Right: using the rescaling of the muon mass using $f_{\pi}$. }
\vskip 0.1in
\renewcommand{\arraystretch}{1.1}
{\footnotesize
\begin{tabular}{l@{\hskip 01em}|@{\hskip 01em}c@{\hskip 01em}c@{\hskip 01em}c@{\hskip 01em}|@{\hskip 01em}c@{\hskip 01em}c@{\hskip 01em}c}
\hline
&
\multicolumn{3}{c|@{\hskip 01em}}{Standard kernel} &
\multicolumn{3}{c@{\hskip 01em}}{Kernel with rescaling using $f_{\pi}$} \\
&
cut 300~MeV &
cut 360~MeV &
no cut &
cut 300~MeV &
cut 360~MeV &
no cut   \\
\hline
Fit Eq.~(\ref{eq:extrap_log}) &
700(22) & 695(19) &
700(18) & 675(14) & 671(11) & 671(10)  \\
Fit Eq.~(\ref{eq:extrap_ys})  &
700(23) & 689(19) &
683(17) & 669(14) & 656(09) & 645(07)  \\
Fit Eq.~(\ref{eq:extrap_1ovy})
& 700(22) &
697(19) & 704(18) &  677(14) & 676(12) & 681(11)  \\
Fit Eq.~(\ref{eq:extrap_ylog})
& 700(22) &
692(19) & 692(17) &
672(14) & 663(10) & 657(08)  \\
\hline
\end{tabular} 

}
\label{tab:fit}
\end{table}

\subsection{The quark-disconnected contribution}

\begin{figure}[t!]
         \includegraphics*[width=0.72\linewidth]{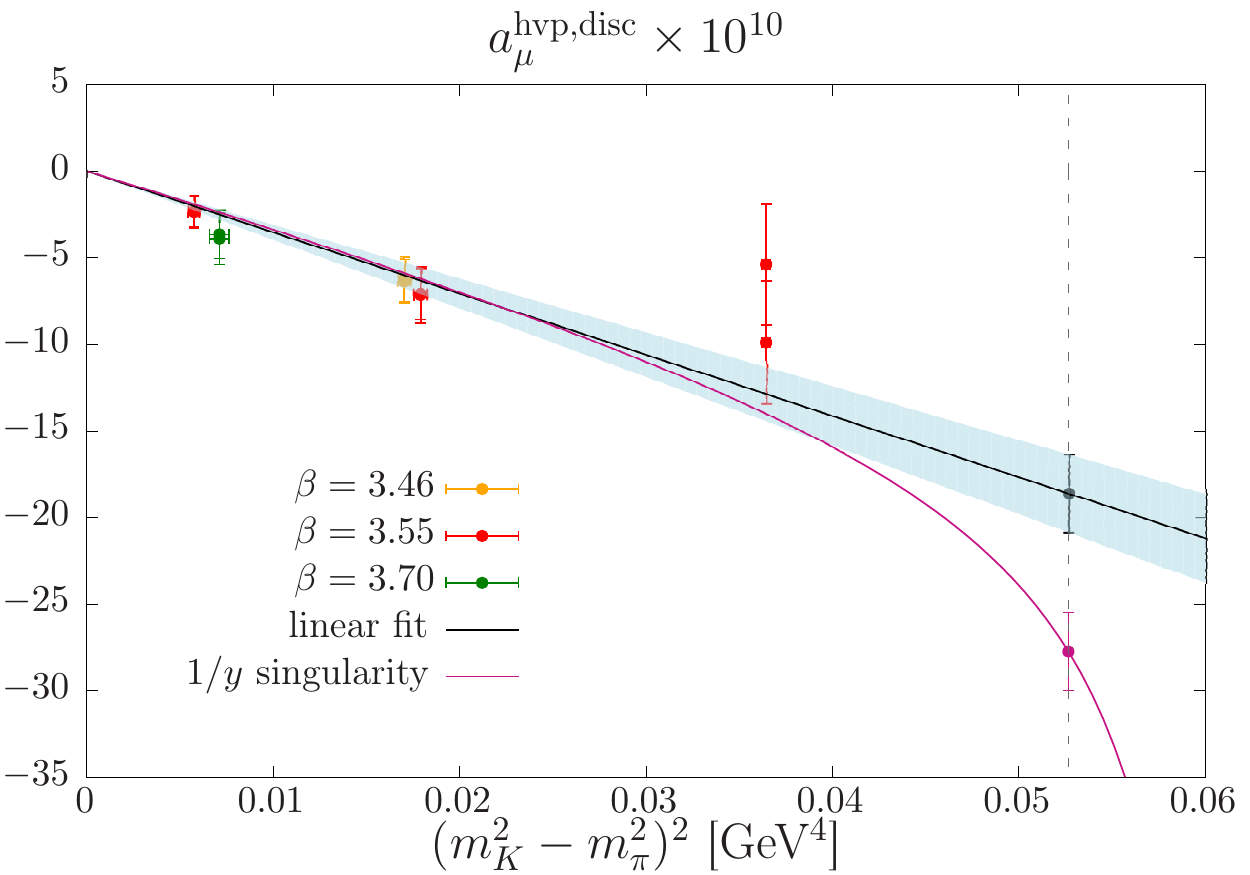}
        \caption{Extrapolation of the disconnected contribution to $\ahvp$ in the SU(3)-breaking variable
$\Delta_2\equiv m_K^2-m_\pi^2$. The data points for the local-local and the local-conserved discretizations are shown.
A linear fit (straight black line), as well as a fit based on ansatz (\ref{eq:ansatzdisc}) are shown.}
        \label{fig:amudisc}
\end{figure}

The quark-disconnected contributions have been computed on a subset of
the gauge ensembles, as described in Section \ref{sec:simpar}. 
 Three ensembles at the same lattice spacing -- N203, N200 and D200 --
allow us to study the chiral behaviour. Two other ensembles, N401
and N302, enable us to constrain the discretization effects. 

The quark-disconnected contribution vanishes exactly for the ensembles generated at the SU(3) symmetric point.
In fact, it is a double zero in the SU(3) breaking combination $(m_s-m_l)$.
Since our ensembles follow a chiral trajectory at fixed bare average quark mass $(2\mql+\mqs)$,
we can consider the values of $a_\mu^{\rm disc}$ as being to a good approximation\footnote{A residual dependence on 
the independent combination $(\frac{1}{2}m_\pi^2 +m_K^2)$ persists at higher orders in the chiral expansion and 
via O($a$) discretization effects.} a function
of the single variable $m_K^2-m_\pi^2$.
The results of all five ensembles are thus displayed in Fig.\ \ref{fig:amudisc}
as a function of $\Delta_2^{\,2}$, where $\Delta_2\equiv m_K^2-m_\pi^2$, since close to the SU(3) symmetric point, 
the dependence of $a_\mu^{\rm disc}$ on $\Delta_2^{\,2}$ is linear. 
We observe that discretization effects are negligible at the current level of
precision. The result of an extrapolation to the physical point $\Delta_2=0.227\,\GeV^2$ assuming a
linear proportionality to $\Delta_2^{\,2}$ is $a_\mu^{{\rm hvp,\,disc}}=-18.6(2.2)\times10^{-10}$.

As discussed below Eq.\ (\ref{eq:amudiscchiral}), the disconnected contribution
has a singular behaviour in the limit $m_\pi\to 0$, closely related to
the corresponding behaviour of the connected light contribution.
Therefore, we consider the possibility that the disconnected contribution contains
a term with precisely the dependence given in Eq.\ (\ref{eq:amudiscchiral}).
In order to make this term consistent with the double zero of the disconnected correlator 
at $\Delta_2=0$, we fix $\hat M^2 \equiv \frac{1}{2}m_\pi^2 + m_K^2$ to its physical value,
express $m_\pi^2$ through the variable $\Delta_2$ and use the ansatz
\be\la{eq:ansatzdisc}
a_\mu^{{\rm hvp,\,disc}}(\Delta_2) = \gamma_8 \Delta_2^{\,2} -\frac{\alpha^2m_\mu^2}{3240\pi^2} \cdot
\frac{3}{2}\Big[ \frac{1}{\hat M^2 - \Delta_2}   - \frac{\Delta_2}{\hat M^4} - \frac{1}{\hat M^2}\Big].
\ee
Fitting the single free parameter $\gamma_8$, we obtain $a_\mu^{{\rm hvp,\,disc}}=-27.7(2.2)\times10^{-10}$.
From Fig.\ \ref{eq:amudiscchiral}, it is clear that both the linear fit in $\Delta_2^{\,2}$ and the one 
based on ansatz (\ref{eq:ansatzdisc}) are consistent with the lattice data.
While a singular chiral behaviour must be present in $a_\mu^{{\rm hvp,\,disc}}$, 
the ansatz (\ref{eq:ansatzdisc}) may lead to an overestimate of this effect.
Therefore, we quote as our final result the average of the linear and the chirally singular fit,
\be\la{eq:amudisc}
a_\mu^{{\rm hvp,\,disc}}= (-23.2 \pm 2.2 \pm 4.5)\times 10^{-10},
\ee
where the first error is statistical and 
the second  is a systematic error associated with 
the extrapolation to the physical point, taken to be the half-distance between the two extrapolated values.

\subsection{The total $\ahvp$ }

In summary, adding up the connected light, strange and charm contributions  as well as the quark-disconnected
contribution, our result for $\ahvp$ in isospin-symmetric QCD at $m_\pi=134.97\,$MeV and $f_\pi=92.4\,\MeV$ is 
\be\la{eq:amuQCD}
\ahvp = (720.0\pm 12.4 \pm 6.8)\times 10^{-10},
\ee
where the first error is statistical and the second is the systematic error.
The latter is dominated by the chiral extrapolation of the light-connected and the disconnected contributions.
The result Eq.\ (\ref{eq:amuQCD}) does not contain any correction for QED or strong isospin-breaking effects.
For now, we do not attempt to include such a correction, but rather add (in quadrature) a systematic uncertainty
of $7.2\times 10^{-10}$  corresponding to a recent lattice calculation of these effects~\cite{Giusti:2019xct}.
This then leads to our final result given in Eq.\ (\ref{eq:final}) below.

\section{Discussion and comparison\label{sec:discussion}}

\begin{figure}[t!]
        \includegraphics*[width=0.99\linewidth]{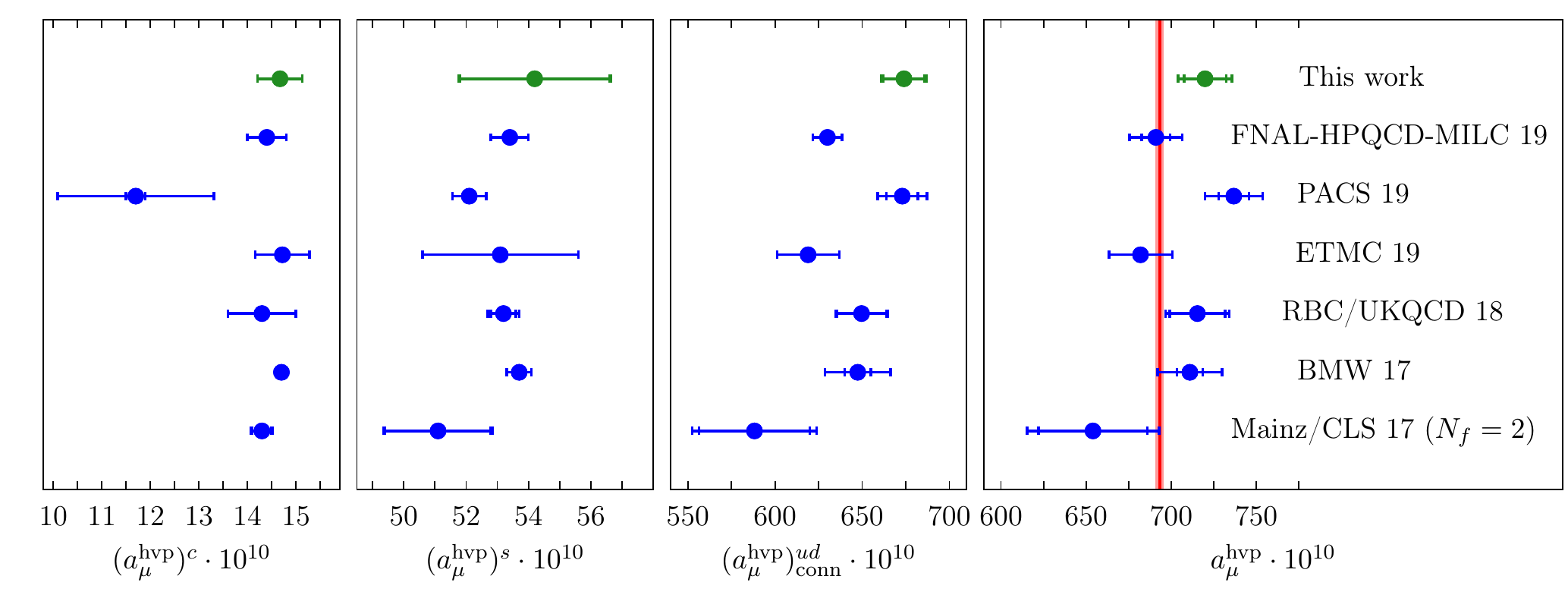}

        \caption{Compilation of lattice results for the connected contributions to
  $\ahvp$ from individual charm, strange and light quarks (left to right). 
  In the rightmost panel,
  the full results, including (where available) the contributions from
  quark-disconnected diagrams and corrections due to isospin-breaking,
  are compared to the phenomenological determination of
  Ref. \cite{Keshavarzi:2018mgv}, represented by the red vertical
  band. Our result is compared to the calculations labelled
  FNAL-HPQCD-MILC\,19 \cite{Chakraborty:2014mwa, Chakraborty:2017tqp,
    Davies:2019efs}, PACS\,19 \cite{Shintani:2019wai}, ETMC\,19
  \cite{Giusti:2017jof, Giusti:2018mdh, Giusti:2019xct}, RBC/UKQCD\,18
  \cite{Blum:2018mom}, BMW\,17 \cite{Borsanyi:2017zdw}, as well as our
  previous calculation in two-flavour QCD \cite{DellaMorte:2017dyu}
  (Mainz/CLS\,17).
}
        \label{fig:amucomp}
\end{figure}

In this paper we have presented a calculation of the hadronic vacuum
polarization contribution to $a_\mu$ based on gauge ensembles with
$N_f=2+1$ flavours of O($a$) improved Wilson quarks. Our final result
is
\begin{equation} \la{eq:final}
 \ahvp=(720.0\pm12.4_{\rm stat}\,\pm9.9_{\rm syst})\cdot10^{-10},
\end{equation}
where the first error is statistical, and the second is an estimate of
the total systematic uncertainty, which also accounts for the fact
that the corrections due to isospin breaking have not been included.
We thus find that the overall error of our determination is 2.2\%. In
Fig.~\ref{fig:amucomp} we compare our results to those of several
other recent lattice calculations \cite{DellaMorte:2017dyu,
Borsanyi:2017zdw, Blum:2018mom, Giusti:2019xct, Shintani:2019wai,
Davies:2019efs}. While our estimate is at the higher end of lattice
results, we note that the direct difference with the result based on
dispersion theory of Ref.~\cite{Keshavarzi:2018mgv} is $26.6\pm16.0$,
which amounts to $\sim1.7$ standard deviations and may signal a slight
tension.

There are several ways in which our result can be improved without
relying on the obvious strategy of adding more ensembles and
increasing the overall statistics. First, we have seen in section
\ref{sec:D200} that the use of detailed spectroscopy information in
the isovector channel is a huge advantage, as it nearly halves the
statistical uncertainty in the estimate for $a_\mu^{{\rm hvp},l}$ on ensemble
D200. This is the result of either constructing the vector correlator
from the energies and overlaps determined via the GEVP or of using
this information in the improved bounding method. Extending these
calculations to more ensembles -- in particular those with physical
and near-physical pion masses -- will boost the statistical accuracy
and reliability significantly.

Second, we have pointed out that it is advantageous to split the
correlator into isovector and isoscalar components according to
Eq.~(\ref{eq:GdecompI}) rather than focussing on separating the
contributions from individual quark flavours. One reason is that the
singular chiral behaviour expected from Eq.~(\ref{eq:amu_chpt2}) is
shared between the light quark connected and the disconnected
contributions. This will help to better constrain the pion mass
dependence of the quark-disconnected contribution, which is often
still obtained from an extrapolation to the physical point from a set
of results at heavier pion masses. The decomposition according to
isospin also gives a better handle on finite-volume effects, which are
partly compensated between the light connected and disconnected
contributions. This is of particular importance, since finite-volume
corrections for the disconnected part of the vector correlator could
be sizeable but, to our knowledge, have not been estimated so far.

The third refinement concerns the determination of isospin-breaking
corrections. We stress again that our final estimate in
Eq.~(\ref{eq:final}) is valid at a well-defined reference point of the
isospin-symmetric theory, given by the mass of the neutral pion in the
continuum limit. The determination of the corrections due to isospin
breaking relies on the definition of an alternative reference point
that is consistent with the effects induced by a non-vanishing mass
splitting among the up and down quarks and the coupling between quarks
and photons. This requires an adjustment of bare parameters and the
re-evaluation of a number of observables that enter the calculation of
$\ahvp$. An account of the status of our activities in this direction
is given in Refs. \cite{Risch:2017xxe,Risch:2018ozp}. In the absence
of a complete evaluation, we have refrained from simply adding results
for the isospin-breaking correction from the literature. Instead, we
have opted for an additional systematic error which is as large as the
correction determined in \cite{Giusti:2019xct}.

As the community awaits the first results from the E989 experiment at
Fermilab, it is remarkable that several collaborations using different
setups and discretizations of the QCD action obtain largely consistent
estimates for $\ahvp$ with overall errors at the level of
2\%. However, the collection of available results does not allow for a
firm conclusion as to whether the phenomenological estimate or the
so-called ``No New Physics'' scenario is confirmed.

\vspace{-0.1in}

\acknowledgments{\noindent
We thank D.\ Djukanovic, T.\ Harris, K.\ Miura, A.\ Nyffeler and A.\ Risch for helpful discussions.
This work is partly supported by the Deutsche Forschungsgemeinschaft (DFG, German Research Foundation) grant HI 2048/1-1
and by the DFG-funded Collaborative Research Centre SFB\,1044 \emph{The low-energy frontier of the Standard Model}.
The Mainz $(g-2)_\mu$ project is also supported by the Cluster of Excellence \emph{Precision Physics, Fundamental Interactions, and Structure of Matter} (PRISMA+ EXC 2118/1) funded by the DFG within the German Excellence Strategy (Project ID 39083149).
Calculations for this project were partly performed on the HPC clusters ``Clover'' and ``HIMster II'' at the Helmholtz-Institut Mainz and ``Mogon II'' at JGU Mainz. M.C.\ thanks A.\ Rago for pointing out Ref.\ \cite{Boyle:2017xcy} 
on how to best exploit the network performance on Mogon II and HIMster II in the early stages of running.
Additional computer time has been allocated through project HMZ21 on the BlueGene  supercomputer system ``JUQUEEN'' at NIC, J\"ulich.
The authors also gratefully acknowledge the Gauss Centre for Supercomputing 
e.V.\ (www.gauss-centre.eu) for funding this project by providing 
computing time on the GCS Supercomputer HAZEL HEN at H\"ochstleistungsrechenzentrum Stuttgart (www.hlrs.de) under project GCS-HQCD.
Our programs use the deflated SAP+GCR solver from the openQCD package~\cite{Luscher:2012av}, as well as the QDP++ library
\cite{Edwards:2004sx}.
We are grateful to our colleagues in the CLS initiative for sharing ensembles.
}

\appendix 

\section{Results for the individual contributions to $\ahvp$ and auxiliary calculations\la{sec:ResTabs}}

\begin{table}[h]
\renewcommand{\arraystretch}{1.2}
\caption{Values of the connected strange and charm contribution to $\ahvp$, in units of $10^{-10}$,
 for the local-local (${\scriptstyle\rm LL}$) and for the local-conserved (${\scriptstyle\rm CL}$) discretizations of the correlation function. 
For the strange, the symmetric lattice derivative is used for the improvement term, while the `away' derivative is used for the charm 
(see section \ref{sec:improvt}).
In addition, the charm hopping parameter is given, as well as the mass-dependent
 renormalization factor $Z_{\rm V}^{(c)}$ for the charm vector current.
The bounding method is used to handle the tail of the correlator, as described in section \ref{sec:conn}.
For the charm contribution, the first error is statistical, the second comes from the tuning of the charm hopping parameter.
The scale-setting uncertainty is not included at this stage.}
\vskip 0.1in
\begin{tabular}{l|@{\hskip 01em}c@{\hskip 01em}c@{\hskip 01em}|@{\hskip 01em}c@{\hskip 01em}c@{\hskip 01em}|@{\hskip 01em}c@{\hskip 01em}c}
\hline
id	&	$a_{\mu}^{({\scriptscriptstyle\rm LL}),s}$	&	$a_{\mu}^{({\scriptscriptstyle\rm CL}),s}$	&	$a_{\mu}^{({\scriptscriptstyle\rm LL}),c}$	&	$a_{\mu}^{({\scriptscriptstyle\rm CL}),c}$	&	$\kappa_{\rm c}$ &	$Z_{\rm V}^{(c)}$ \\
\hline
H101
& 91.09(42)
& 92.47(41)
& 23.909(31)(56)
& 12.541(21)(31)
& 0.122897(18)  & 1.20324(11)(25) \\ 
H102
& 81.10(37) 
& 82.47(36)
& 24.088(20)(81)
& 12.709(20)(46) & 0.123041(26)
& 1.19743(08)(20)  \\ 
U101$^*$
& 71.08(44)
& 72.40(44)
& 24.510(28)(61) & 13.051(19)(35)
& 0.123244(19)  & 1.18964(08)(15) \\ 
H105$^*$
& 71.94(31)
& 73.27(31)
& 24.437(42)(61) & 12.996(29)(35)
& 0.123244(19)  & 1.18964(08)(15) \\ 
N101
& 72.14(23) 
& 73.50(23)
& 24.414(57)(61)
& 12.996(38)(35)
& 0.123244(19)  & 1.18964(08)(15) \\ 
C101
& 68.37(18)
& 69.70(18)
& 24.580(43)(39)
& 13.140(29)(22)
& 0.123361(12)  & 1.18500(05)(10) \\ 
\hline
B450
& 87.27(49)
& 88.41(48) 
& 21.793(24)(70)
& 12.668(18)(43)
& 0.125095(22)  & 1.12972(06)(16) \\ 
S400
& 77.72(34)
& 78.78(34)
& 21.808(27)(64) &
12.919(20)(41)  & 0.125252(20) & 1.11159(13)(18)
\\
N401
& 70.80(20)
& 71.81(20) 
& 22.390(39)(50)
& 13.248(29)(32)
& 0.125439(15)  & 1.11412(04)(19) \\ 
\hline
H200$^*$
& 83.02(87)
& 83.69(86)
& 20.018(37)(56)
& 13.248(28)(40)
& 0.127579(16)  & 1.04843(03)(19) \\ 
N202
& 88.83(71)
& 89.61(69)
& 20.052(39)(56)
& 13.280(30)(40)
& 0.127579(16)  & 1.04843(03)(53) \\ 
N203
& 76.17(37)
& 76.91(36)
& 19.969(30)(39)
& 13.252(22)(28)
& 0.127714(11)  & 1.04534(03)(19)  \\ 
N200 
& 67.90(23)
& 68.60(22)
& 20.323(42)(26)
& 13.577(32)(18)
& 0.127858(07)  & 1.04012(03)(13) \\ 
D200 
& 62.20(15)
& 62.93(15)
& 20.677(39)(22)
& 13.895(30)(16)
& 0.127986(06)  & 1.03587(04)(11) \\ 
E250 
& 59.64(14)
& 60.36(14)
& 20.798(02)(21) &
14.027(02)(15) & 0.128052(05)
& 1.03310(01)(10)  \\
\hline
N300
& 81.03(69)
& 81.44(68)
& 17.367(34)(73)
& 13.159(29)(58)
& 0.130099(18)  & 0.97722(03)(12) \\ 
N302    
& 70.11(31)
& 70.53(30)
& 17.839(28)(38) &
13.606(23)(30)  & 0.130247(09) & 0.97241(03)(10) \\ 
J303
& 62.87(21)
& 63.27(20)
& 17.931(49)(31)
& 13.870(40)(27)
& 0.130362(09)  & 0.96037(10)(19) \\ 
\hline
\end{tabular} 

        \label{tab:resultsSC}
\end{table}

\begin{table}[h]
\renewcommand{\arraystretch}{1.15}
\caption{Results for $a_\mu^{\rm hvp,disc}$, with and without rescaling the muon mass.}
\vskip 0.1in
\begin{tabular}{l|@{\hskip 01em}c@{\hskip 01em}c@{\hskip 01em}|@{\hskip 01em}c@{\hskip 01em}c}
\hline
&
\multicolumn{2}{c|@{\hskip 01em}}{No rescaling} 
& \multicolumn{2}{c}{With rescaling} \\
\hline
id
& $a_{\mu}^{({\scriptscriptstyle\rm LL}),l}$
& $a_{\mu}^{({\scriptscriptstyle\rm CL}),l}$       &
$a_{\mu}^{({\scriptscriptstyle\rm LL}),l}$  & $a_{\mu}^{({\scriptscriptstyle\rm CL}),l}$
\\
\hline
N401
& -5.41(1.25) &  -5.50(1.21)
& -5.76(1.30) & -5.85(1.26) \\
\hline
N203
& -1.60(0.77) & -1.65(0.80)
& -1.85(0.88)  & -1.91(0.92) \\
N200
& -5.50(1.32) &  -5.55(1.46)
& -5.99(1.46) & -6.05(1.61) \\
D200 
& -3.79(3.51) & -8.32(3.57) & -3.78(3.50)
& -8.30(3.56) \\
\hline
N302
& -1.55(1.16) & -1.77(1.23)
& -1.84(1.39) & -2.10(1.47) \\
\hline
\end{tabular} 
\label{tab:disc}
\end{table}

\newpage

\begin{table}[h!]
        \caption{Values of the connected light contribution to $\ahvp$, in units of $10^{-10}$, 
 for the local-local (${\scriptstyle\rm LL}$) and for the local-conserved (${\scriptstyle\rm CL}$) 
discretizations of the correlation function, as described in the main text. 
The symmetric lattice derivative of the improvement term is used. 
In addition, the pion mass and the pion decay constant are given, some of them taken from Ref.~\cite{Gerardin:2019vio}. 
No FSE correction has been applied on any data in the table.
The treatment of the long-distance part of the correlator is described in section \ref{sec:conn}.
In the ``No rescaling'' columns, the scale-setting error has not been included. In the ``With rescaling'' columns, the 
statistical fluctuations of the $f_\pi$ determination are taken into account. }
\vskip 0.1in

\begin{tabular}{l|@{\hskip 01em}c@{\hskip 01em}c@{\hskip 01em}|@{\hskip 01em}c@{\hskip 01em}c@{\hskip 01em}|@{\hskip 01em}c@{\hskip 01em}c}
\hline
&
& &
\multicolumn{2}{c|@{\hskip 01em}}{No rescaling} 
& \multicolumn{2}{c}{With rescaling} \\
\hline
id		&	$am_{\pi}$	&	$a\sqrt{2}f_{\pi}$	&	 $a_{\mu}^{({\scriptscriptstyle\rm LL}),l}$	&	$a_{\mu}^{({\scriptscriptstyle\rm CL}),l}$       &	$a_{\mu}^{({\scriptscriptstyle\rm LL}),l}$	&	$a_{\mu}^{({\scriptscriptstyle\rm CL}),l}$	\\
\hline
H101
& 0.1818(07)  & 0.06458(29) & 455.5(2.0)
& 462.4(2.0)  & 571.9(4.4) & 580.5(4.5) \\ 
H102
& 0.1547(08)  & 0.06151(27) & 495.0(3.5)  & 501.9(3.4)
& 572.2(4.7) &  580.2(4.7) \\ 
H105$^*$
& 0.1224(11)  & 0.05802(39) & 548.7(5.9)
& 555.4(6.0) & 576.4(8.0) & 583.4(8.1) \\ 
N101
& 0.1217(06)  & 0.05832(30) & 562.8(4.3)
& 568.9(4.2)  & 589.8(6.8) & 596.4(6.7) \\
C101
& 0.0967(08)  & 0.05535(37) & 625.0(6.7)
& 633.3(7.5)  & 612.9(6.9) & 621.9(7.3) \\ 
\hline
B450
& 0.1608(05)  & 0.05750(27)  & 436.3(2.4) 
& 442.0(2.4)  & 556.8(4.7) & 564.3(4.8) \\
S400
& 0.1357(05)  & 0.05463(21) & 481.4(3.8) & 486.4(3.8)
& 564.3(5.2) & 570.2(5.2) \\
N401
& 0.1104(06)  & 0.05324(17) & 543.8(4.2) & 548.9(4.1)
& 581.5(5.3) & 587.0(5.2) \\
\hline
H200$^*$
& 0.1362(07)  & 0.04805(27) & 413.6(4.5) & 417.8(4.3)
& 532.0(5.5) & 537.5(5.3) \\
N202
& 0.1335(05)  & 0.04884(18) & 441.8(3.0) & 445.9(2.9)
& 567.9(4.4) &  573.2(4.3)\\
N203
& 0.1126(04)  & 0.04699(16) & 479.1(3.3) & 483.2(3.3)
& 564.6(5.0) &  569.4(5.1) \\
N200 
& 0.0920(05)  & 0.04454(18) & 520.0(5.3) & 524.4(5.0)
& 571.8(6.1) &  576.6(5.8) \\
D200 
& 0.0649(04)  & 0.04254(18) & 600.3(5.0) & 604.3(5.9)
& 598.6(6.3) &  602.4(6.3) \\
E250
& 0.0422(04)
& 0.04089(19)  & 735.1(14.6) & 726.3(14.8)
&  679.7(14.8) & 671.6(14.9) \\
\hline
N300
& 0.1063(04)  & 0.03811(13) & 404.1(3.4) & 406.0(3.3)
& 540.6(5.5) &  543.2(5.4) \\
N302    
& 0.0872(04)  & 0.03570(19) & 436.6(4.4) & 438.8(4.4)
& 535.4(6.5) & 538.1(6.5) \\
J303
& 0.0651(03)  & 0.03412(14) & 526.6(7.4) & 527.4(7.2)
& 577.5(8.5) & 578.3(8.4) \\
\hline
\end{tabular} 
                
        \label{tab:resultsL}
\end{table}

\begin{table}[t]
\caption{Estimates of the finite-size effects $\Delta \ahvp \equiv \ahvp(L=\infty) - \ahvp(L)=\Delta a_{\mu}^{<}(t_i)+
\Delta a_{\mu}^{>}(t_i)$ on the isovector contribution $a_\mu^{{\rm hvp},I=1}$ in the TMR in units of $10^{-10}$. We used the value $t_i = (m_{\pi} L/4)^2 / m_{\pi}$, as in~\cite{DellaMorte:2017dyu}, to which we refer for unexplained notation. The parameters of the GS model are obtained by fitting the tail of the correlation function using the L\"uscher formalism. When there are two lines, the second line corresponds to the GS parameters  extracted from a direct lattice calculation of the timelike pion form factor~\cite{Andersen:2018mau}. 
The FSE effects are given both with and without rescaling of the muon mass via $f_\pi$.
}
\vskip 0.1in
\renewcommand{\arraystretch}{1.1}
\begin{tabular}{lc@{\hskip 01em}|@{\hskip 01em}c@{\hskip 01em}c@{\hskip 01em}|@{\hskip 01em}c@{\hskip 01em}c@{\hskip 01em}|@{\hskip 01em}c@{\hskip 01em}c@{\hskip 01em}}
\hline
&
& \multicolumn{2}{c|@{\hskip 01em}}{GS parameters}
& \multicolumn{2}{c|@{\hskip 01em}}{No rescaling} &
\multicolumn{2}{c}{With rescaling}  \\
%-----------------------------------------------------------------------------------------------------------------------------------------------------------------------------------
id
& $t_i~$[fm]
& $m_{\rho}/m_{\pi}$
& $g_{\rho\pi\pi}$
& $\Delta a_{\mu}^{<}(t_i)$
& $\Delta a_{\mu}^{>}(t_i)$ 
&
$\Delta a_{\mu}^{<}(t_i)$
& $\Delta a_{\mu}^{>}(t_i)$ \\
%-----------------------------------------------------------------------------------------------------------------------------------------------------------------------------------
\hline
H101
& 1.04
& 2.101(103)
& 4.60(42)
& 0.29
& \ 2.0(0.4)
& 0.37
& \ 2.4(0.5)
\\
H102
& 0.86
& 2.307(07)
& 4.88(02)
& 0.40
& \ 5.6(0.1)
& 0.47
& \ 6.3(0.1)
\\
U101$^*$
& 0.35
& 2.698(290)
& 5.23(52)
& 0.11
& 49.0(10.8)
& 0.12
& 51.4(11.3)
\\
H105$^*$
& 0.69
& 2.743(110)
& 5.01(24)
&     0.46
& 16.6(1.4)
& 0.48
& 17.3(1.5)
\\
N101
& 1.55
& 2.792(10)
& 4.93(04)
& 0.61
& \ 1.8(0.1)
& 0.64
& \ 1.9(0.1)
\\ 
C101
& 1.21
& 3.362(73)
& 4.92(12)
& 0.92
& \ 8.1(0.3)
& 0.90
& \ 8.0(0.3)
\\ 
&
1.21 &
3.395(26) &
5.67(17) &
0.92 &
\ 7.9(0.2) &
0.90 &
\ 7.8(0.2) \\
\hline
B450
& 0.76
& 2.100(10)
& 4.86(04)
& 0.27
& \ 4.4(0.1)
& 0.35
& \ 5.6(0.1)
\\ 
S400
& 0.69
& 2.299(41)
& 5.01(14)
& 0.38
& 11.0(0.5)
& 0.45
& 12.8(0.6)
\\ 
N401
& 1.22
& 2.716(25)
& 5.08(06)
& 0.60
& \ 3.7(0.1)
& 0.65
& \ 3.9(0.1)
\\ 
&
1.22 &
2.717(16) &
5.84(17) &
0.60 &
\ 3.6(0.1) &
0.65 &
\ 3.8(0.1) \\ 
\hline
H200$^*$
& 0.58
& 2.095(02)
& 4.86(07)
& 0.27
& 10.5(0.2)
& 0.40
& 13.3(0.2)
\\ 
N202
& 1.22
& 2.016(04)
& 5.21(07) 
& 0.23
& \ 1.1(0.1)
& 0.30
& \ 1.4(0.2)
\\ 
N203
& 1.03
& 2.398(13)
& 4.91(05)
& 0.40
& \ 3.2(0.1)
& 0.47
& \ 3.7(0.1)
\\ 
&
1.03 &
2.382(11) &
6.03(13) &
0.40 &
\ 3.1(0.1) &
0.47 &
\ 3.6(0.1) \\
N200 
& 0.84
& 2.833(20)
& 4.98(05)
& 0.50
& \ 9.2(0.2)
& 0.60
& \ 9.3(0.8)
\\ 
&
0.85 &
2.733(16) &
5.94(10) &
0.49 &
\ 9.6(0.2) &
0.55 &
10.6(0.2) \\ 
D200
& 1.09
& 3.737(72)
& 5.26(10)
& 1.00
& 13.4(0.4)
& 0.99
& 13.3(0.4)
\\ 
&
1.09 &
3.877(34) &
6.16(19) &
1.00 &
12.8(0.2) & 
0.99 &
12.7(0.2) \\ 
E250 
& 1.54
& 5.270(42) 
& 5.59(03)
& 1.88
& 20.7(0.2)
& 1.74
& 19.4(0.2)
\\ 
&
1.54 &
5.955(84) &
6.06(21) &
1.88 &
19.1(0.2) &
1.74 &
17.9(0.2) \\ 
\hline
N300
& 0.75
& 2.100(01)
& 4.89(08)
& 0.27
& \ 4.6(0.2)
& 0.36
& \ 6.1(0.3)
\\
N302    
& 0.65
& 2.301(07)
& 5.59(08)
& 0.36
& 13.0(0.3)
& 0.45
& 15.8(0.3)
\\ 
J303
& 0.85
& 2.993(02)
& 5.17(03)
& 0.61
& 12.4(0.1)
& 0.67
& 13.4(0.1)
\\ 
&
0.85 &
3.090(24) &
6.33(16) &
0.61 &
11.4(0.5) &
0.67 &
12.4(0.5) \\ 
\hline
 \end{tabular} 
\label{tab:FSE}
\end{table}

\begin{table}
\caption{Results of spectroscopy calculations in the isovector vector channel for the parameters
$m_\rho$ and $g_{\rho\pi\pi}$ of the $\rho$ meson. See Eqs.\ (\ref{eq:effrange}) and (\ref{eq:Gamrho}).
For ensemble J303, the data from~\cite{Andersen:2018mau} has been rebinned and analyzed using the jackknife method.}
\vskip 0.1in

        \centering
        \begin{tabular}{c@{~~~}c@{~~~} c@{~~~} c@{~~~} l}
%                \toprule \toprule
\hline
                & $a m_\rho$ & $m_\rho / m_\pi$ & $g_{\rho\pi\pi}$ &  source \\
%                \midrule
\hline
%                 D101 & 0.3285(14) & 3.366(15) & 6.19(10) & \texttt{1808.05007}\\
                C101 & 0.3327(23) & 3.395(26) & 5.67(17) & \cite{Andersen:2018mau}\\
\hline
%                \midrule
                N401 & 0.2989(16) & 2.717(16) & 5.84(17) & \cite{Andersen:2018mau}\\
 \hline
%               \midrule
                N203 & 0.2682(13) & 2.382(11) & 6.03(13) & New data\\
                N200 & 0.2522(13) & 2.733(16) & 5.94(10) & \cite{Andersen:2018mau}\\
                D200 & 0.2501(12) & 3.839(18) & 6.065(92) & New data \\
 \hline
%               \midrule
                J303 & 0.2020(15) & 3.090(24) & 6.33(16) & \cite{Andersen:2018mau} \\
 \hline
%               \bottomrule
        \end{tabular}
\la{tab:spectro}
\end{table}

\newpage

\section{Determination of the charm-quark hopping parameter\label{sec:apda}}

We list in Table \ref{tab:charmTableLong} the values of the $c\bar s$
pseudoscalar meson masses determined for different values of the charm hopping parameter.
The `physical' value of the charm-quark hopping parameter  is determined by the condition
that the $c\bar s$ meson mass match the physical value of the $D_s$ meson mass, $m_{D_s}=1972\,$MeV.

\begin{table}[h!]

\begin{tabular}{l@{\hskip 02em}l@{\hskip 03em}l@{\hskip 03em}llll}
\hline
id		&				&	Interpolation	&	 	\multicolumn{3}{c}{Simulations}		\\
\hline
H101	&	$\kappa$		&	0.122897(18)	& 0.12320 & 0.12290 & 0.12260 & 0.12230\\  
		&	$am_{D_s}$	&	0.8615	& 0.8513(6) & 0.8614(6) & 0.8714(6) & 0.8813(6) \\    
\hline
H102	&	$\kappa$		&	0.123041(26)	& 0.12330 & 0.12300 & 0.12270 & 0.12240\\ 
		&	$am_{D_s}$	&	0.8615		& 0.8528(9) & 0.8629(9) &	 0.8729(9) & 0.8828(9)\\
\hline	
N101 	&	$\kappa$		&	0.123244(19)	& 0.12360 & 0.12340 & 0.12320 & 0.12300	 \\
		&	$am_{D_s}$	&	0.8615		& 0.8495(6) & 0.8563(6) & 0.8630(6) & 0.8697(6) \\
\hline
C101	&	$\kappa$		&	0.123361(12)	& 0.12400 & 0.12370 & 0.12340 & 0.12310\\
		&	$am_{D_s}$	&	0.8615		& 0.8467(4) & 0.8534(4) & 0.8602(4) & 0..8669(4)	 \\
\hline
\hline
B450		&	$\kappa$		&	0.125095(22)	& 0.12530 & 0.12510 & 0.12490 & 0.12470 \\
		&	$am_{D_s}$	&	0.7615		& 0.7543(8) & 0.7614(8) & 0.7683(8) & 0.7752(8) \\
\hline
S400		&	$\kappa$		&	0.125252(20)	& 0.12570 & 0.12550 & 0.12530 & 0.12510 \\
		&	$am_{D_s}$	&	0.7615		& 0.7457(7) & 0.7528(7) & 0.7599(7) & 0.7669(7)	 \\
\hline
N401	&	$\kappa$		&	0.125439(15)	& 0.12570 & 0.12550 & 0.12530 & 0.12510 \\
		&	$am_{D_s}$	&	0.7615		& 0.7523(5) & 0.7594(5) & 0.7564(5) & 0.7734(5)	 \\            
\hline
\hline
N202 	&	$\kappa$		&	0.127579(16) 	& 0.12775 & 0.12765 & 0.12755 & 0.12745	\\
		&	$am_{D_s}$	&	0.6410		& 0.6347(6) & 0.6384(6) &	0.6421(6)	& 0.6458(6)	\\  
\hline
N203	&	$\kappa$		&	0.127714(11)	& 0.12790 & 0.12770 & 0.12750 & 0.12730 \\
		&	$am_{D_s}$	&	0.6410		& 0.6341(4) & 0.6416(4) & 0.6490(4) &	0.6563(4) \\
\hline      
N200  	&	$\kappa$		&	0.127858(07)	& 0.12810 & 0.12790 & 0.12770 & 0.12750	\\
		&	$am_{D_s}$	&	0.6410		& 0.6320(3) & 0.6395(3) &	0.6469(3) & 0.6543(3) \\
\hline
{ D200} 	&	$\kappa$	&	0.127986(06)	& 0.12820 & 0.12810 & 0.12800 & 0.12790 \\
		&	$am_{D_s}$	&	0.6410		& 0.6330(2) & 0.6367(2) & 0.6405(2) & 0.6442(2)	 \\
\hline
E250 	&	$\kappa$		&	0.128052(05)	& 0.12830	& 0.12810 & 0.12790 & 0.12770 \\
		&	$am_{D_s}$	&	0.6410		& 0.6317(2) & 0.6392(2) & 0.6467(2) &	0.6541(2) \\
\hline
\hline
N300	&	$\kappa$		&	0.130099(18)	& 0.12970	& 0.13000 & 0.13030 & 0.13060 \\
		&	$am_{D_s}$	&	0.4969		& 0.5126(7) & 0.5008(7) &  0.4888(7) & 0.4766(7)\\
\hline
N302    	&	$\kappa$		&	0.130247(09)	& 0.13000 & 0.13030 & 0.13060 & 0.13090 \\
		&	$am_{D_s}$	&	0.4969		& 0.5066(3) & 0.4947(3) & 0.4827(3) &	0.4704(3)\\
\hline
J303		&	$\kappa$		&	0.130362(09) 	& 0.13020 & 0.13040 &  0.13060 & 0.13080 \\
		&	$am_{D_s}$	&	0.4969		& 0.5033(4) & 0.4954(4) & 0.4873(4) & 0.4792(3)	 \\
\hline
\end{tabular} 
              
 \caption{Values of the $c\bar s$ pseudoscalar meson masses determined for different values of the charm hopping parameter,
and its interpolation to the physical $D_s$ meson mass.}
\label{tab:charmTableLong}   
\end{table}

\bibliography{/Users/harvey/BIBLIO/viscobib}

\begin{thebibliography}{64}
\expandafter\ifx\csname natexlab\endcsname\relax\def\natexlab#1{#1}\fi
\expandafter\ifx\csname bibnamefont\endcsname\relax
  \def\bibnamefont#1{#1}\fi
\expandafter\ifx\csname bibfnamefont\endcsname\relax
  \def\bibfnamefont#1{#1}\fi
\expandafter\ifx\csname citenamefont\endcsname\relax
  \def\citenamefont#1{#1}\fi
\expandafter\ifx\csname url\endcsname\relax
  \def\url#1{\texttt{#1}}\fi
\expandafter\ifx\csname urlprefix\endcsname\relax\def\urlprefix{URL }\fi
\providecommand{\bibinfo}[2]{#2}
\providecommand{\eprint}[2][]{\url{#2}}

\bibitem[{\citenamefont{Bennett et~al.}(2006)}]{Bennett:2006fi}
\bibinfo{author}{\bibfnamefont{G.~W.} \bibnamefont{Bennett}}
  \bibnamefont{et~al.} (\bibinfo{collaboration}{Muon g-2}),
  \bibinfo{journal}{Phys. Rev.} \textbf{\bibinfo{volume}{D73}},
  \bibinfo{pages}{072003} (\bibinfo{year}{2006}), \eprint{hep-ex/0602035}.

\bibitem[{\citenamefont{Jegerlehner and Nyffeler}(2009)}]{Jegerlehner:2009ry}
\bibinfo{author}{\bibfnamefont{F.}~\bibnamefont{Jegerlehner}} \bibnamefont{and}
  \bibinfo{author}{\bibfnamefont{A.}~\bibnamefont{Nyffeler}},
  \bibinfo{journal}{Phys.Rept.} \textbf{\bibinfo{volume}{477}},
  \bibinfo{pages}{1} (\bibinfo{year}{2009}), \eprint{0902.3360}.

\bibitem[{\citenamefont{Blum et~al.}(2013)\citenamefont{Blum, Denig,
  Logashenko, de~Rafael, Lee~Roberts et~al.}}]{Blum:2013xva}
\bibinfo{author}{\bibfnamefont{T.}~\bibnamefont{Blum}},
  \bibinfo{author}{\bibfnamefont{A.}~\bibnamefont{Denig}},
  \bibinfo{author}{\bibfnamefont{I.}~\bibnamefont{Logashenko}},
  \bibinfo{author}{\bibfnamefont{E.}~\bibnamefont{de~Rafael}},
  \bibinfo{author}{\bibfnamefont{B.}~\bibnamefont{Lee~Roberts}},
  \bibnamefont{et~al.} (\bibinfo{year}{2013}), \eprint{1311.2198}.

\bibitem[{\citenamefont{Jegerlehner}(2017)}]{Jegerlehner:2017gek}
\bibinfo{author}{\bibfnamefont{F.}~\bibnamefont{Jegerlehner}},
  \emph{\bibinfo{title}{{The Anomalous Magnetic Moment of the Muon}}}
  (\bibinfo{publisher}{Springer Tracts Mod. Phys., vol. 274},
  \bibinfo{year}{2017}).

\bibitem[{\citenamefont{Grange et~al.}(2015)}]{Grange:2015fou}
\bibinfo{author}{\bibfnamefont{J.}~\bibnamefont{Grange}} \bibnamefont{et~al.}
  (\bibinfo{collaboration}{Muon g-2}) (\bibinfo{year}{2015}),
  \eprint{1501.06858}.

\bibitem[{\citenamefont{Mibe}(2011)}]{Mibe:2011zz}
\bibinfo{author}{\bibfnamefont{T.}~\bibnamefont{Mibe}}
  (\bibinfo{collaboration}{J-PARC g-2}), \bibinfo{journal}{Nucl. Phys. Proc.
  Suppl.} \textbf{\bibinfo{volume}{218}}, \bibinfo{pages}{242}
  (\bibinfo{year}{2011}).

\bibitem[{\citenamefont{Meyer and Wittig}(2019)}]{Meyer:2018til}
\bibinfo{author}{\bibfnamefont{H.~B.} \bibnamefont{Meyer}} \bibnamefont{and}
  \bibinfo{author}{\bibfnamefont{H.}~\bibnamefont{Wittig}},
  \bibinfo{journal}{Prog. Part. Nucl. Phys.} \textbf{\bibinfo{volume}{104}},
  \bibinfo{pages}{46} (\bibinfo{year}{2019}), \eprint{1807.09370}.

\bibitem[{\citenamefont{Blum et~al.}(2017)\citenamefont{Blum, Christ, Hayakawa,
  Izubuchi, Jin, Jung, and Lehner}}]{Blum:2016lnc}
\bibinfo{author}{\bibfnamefont{T.}~\bibnamefont{Blum}},
  \bibinfo{author}{\bibfnamefont{N.}~\bibnamefont{Christ}},
  \bibinfo{author}{\bibfnamefont{M.}~\bibnamefont{Hayakawa}},
  \bibinfo{author}{\bibfnamefont{T.}~\bibnamefont{Izubuchi}},
  \bibinfo{author}{\bibfnamefont{L.}~\bibnamefont{Jin}},
  \bibinfo{author}{\bibfnamefont{C.}~\bibnamefont{Jung}}, \bibnamefont{and}
  \bibinfo{author}{\bibfnamefont{C.}~\bibnamefont{Lehner}},
  \bibinfo{journal}{Phys. Rev. Lett.} \textbf{\bibinfo{volume}{118}},
  \bibinfo{pages}{022005} (\bibinfo{year}{2017}), \eprint{1610.04603}.

\bibitem[{\citenamefont{Asmussen et~al.}(2018)\citenamefont{Asmussen,
  G{\'e}rardin, Nyffeler, and Meyer}}]{Asmussen:2018oip}
\bibinfo{author}{\bibfnamefont{N.}~\bibnamefont{Asmussen}},
  \bibinfo{author}{\bibfnamefont{A.}~\bibnamefont{G{\'e}rardin}},
  \bibinfo{author}{\bibfnamefont{A.}~\bibnamefont{Nyffeler}}, \bibnamefont{and}
  \bibinfo{author}{\bibfnamefont{H.~B.} \bibnamefont{Meyer}}, in
  \emph{\bibinfo{booktitle}{{15th International Workshop on Tau Lepton Physics
  (TAU2018) Amsterdam, Netherlands, September 24-28, 2018}}}
  (\bibinfo{year}{2018}), \eprint{1811.08320}.

\bibitem[{\citenamefont{Colangelo et~al.}(2018)\citenamefont{Colangelo,
  Hoferichter, Procura, and Stoffer}}]{Colangelo:2017urn}
\bibinfo{author}{\bibfnamefont{G.}~\bibnamefont{Colangelo}},
  \bibinfo{author}{\bibfnamefont{M.}~\bibnamefont{Hoferichter}},
  \bibinfo{author}{\bibfnamefont{M.}~\bibnamefont{Procura}}, \bibnamefont{and}
  \bibinfo{author}{\bibfnamefont{P.}~\bibnamefont{Stoffer}},
  \bibinfo{journal}{EPJ Web Conf.} \textbf{\bibinfo{volume}{175}},
  \bibinfo{pages}{01025} (\bibinfo{year}{2018}), \eprint{1711.00281}.

\bibitem[{\citenamefont{Bruno et~al.}(2015)}]{Bruno:2014jqa}
\bibinfo{author}{\bibfnamefont{M.}~\bibnamefont{Bruno}} \bibnamefont{et~al.},
  \bibinfo{journal}{JHEP} \textbf{\bibinfo{volume}{02}}, \bibinfo{pages}{043}
  (\bibinfo{year}{2015}), \eprint{1411.3982}.

\bibitem[{\citenamefont{Bali et~al.}(2016{\natexlab{a}})\citenamefont{Bali,
  Scholz, Simeth, and S{\"o}ldner}}]{Bali:2016umi}
\bibinfo{author}{\bibfnamefont{G.~S.} \bibnamefont{Bali}},
  \bibinfo{author}{\bibfnamefont{E.~E.} \bibnamefont{Scholz}},
  \bibinfo{author}{\bibfnamefont{J.}~\bibnamefont{Simeth}}, \bibnamefont{and}
  \bibinfo{author}{\bibfnamefont{W.}~\bibnamefont{S{\"o}ldner}}
  (\bibinfo{collaboration}{RQCD}), \bibinfo{journal}{Phys. Rev.}
  \textbf{\bibinfo{volume}{D94}}, \bibinfo{pages}{074501}
  (\bibinfo{year}{2016}{\natexlab{a}}), \eprint{1606.09039}.

\bibitem[{\citenamefont{Mohler et~al.}(2018)\citenamefont{Mohler, Schaefer, and
  Simeth}}]{Mohler:2017wnb}
\bibinfo{author}{\bibfnamefont{D.}~\bibnamefont{Mohler}},
  \bibinfo{author}{\bibfnamefont{S.}~\bibnamefont{Schaefer}}, \bibnamefont{and}
  \bibinfo{author}{\bibfnamefont{J.}~\bibnamefont{Simeth}},
  \bibinfo{journal}{EPJ Web Conf.} \textbf{\bibinfo{volume}{175}},
  \bibinfo{pages}{02010} (\bibinfo{year}{2018}), \eprint{1712.04884}.

\bibitem[{\citenamefont{Risch and Wittig}(2018{\natexlab{a}})}]{Risch:2017xxe}
\bibinfo{author}{\bibfnamefont{A.}~\bibnamefont{Risch}} \bibnamefont{and}
  \bibinfo{author}{\bibfnamefont{H.}~\bibnamefont{Wittig}},
  \bibinfo{journal}{EPJ Web Conf.} \textbf{\bibinfo{volume}{175}},
  \bibinfo{pages}{14019} (\bibinfo{year}{2018}{\natexlab{a}}),
  \eprint{1710.06801}.

\bibitem[{\citenamefont{Risch and Wittig}(2018{\natexlab{b}})}]{Risch:2018ozp}
\bibinfo{author}{\bibfnamefont{A.}~\bibnamefont{Risch}} \bibnamefont{and}
  \bibinfo{author}{\bibfnamefont{H.}~\bibnamefont{Wittig}}, in
  \emph{\bibinfo{booktitle}{{36th International Symposium on Lattice Field
  Theory (Lattice 2018) East Lansing, MI, United States, July 22-28, 2018}}}
  (\bibinfo{year}{2018}{\natexlab{b}}), \eprint{1811.00895}.

\bibitem[{\citenamefont{Blum}(2003)}]{Blum:2002ii}
\bibinfo{author}{\bibfnamefont{T.}~\bibnamefont{Blum}},
  \bibinfo{journal}{Phys.Rev.Lett.} \textbf{\bibinfo{volume}{91}},
  \bibinfo{pages}{052001} (\bibinfo{year}{2003}), \eprint{hep-lat/0212018}.

\bibitem[{\citenamefont{Bernecker and Meyer}(2011)}]{Bernecker:2011gh}
\bibinfo{author}{\bibfnamefont{D.}~\bibnamefont{Bernecker}} \bibnamefont{and}
  \bibinfo{author}{\bibfnamefont{H.~B.} \bibnamefont{Meyer}},
  \bibinfo{journal}{Eur.Phys.J.} \textbf{\bibinfo{volume}{A47}},
  \bibinfo{pages}{148} (\bibinfo{year}{2011}), \eprint{1107.4388}.

\bibitem[{\citenamefont{L{\"u}scher}(1991)}]{Luscher:1991cf}
\bibinfo{author}{\bibfnamefont{M.}~\bibnamefont{L{\"u}scher}},
  \bibinfo{journal}{Nucl. Phys.} \textbf{\bibinfo{volume}{B364}},
  \bibinfo{pages}{237} (\bibinfo{year}{1991}).

\bibitem[{\citenamefont{Meyer}(2011)}]{Meyer:2011um}
\bibinfo{author}{\bibfnamefont{H.~B.} \bibnamefont{Meyer}},
  \bibinfo{journal}{Phys.Rev.Lett.} \textbf{\bibinfo{volume}{107}},
  \bibinfo{pages}{072002} (\bibinfo{year}{2011}), \eprint{1105.1892}.

\bibitem[{\citenamefont{Della~Morte et~al.}(2017)\citenamefont{Della~Morte,
  Francis, G{\"u}lpers, Herdo{\'i}za, von Hippel, Horch, J{\"a}ger, Meyer,
  Nyffeler, and Wittig}}]{DellaMorte:2017dyu}
\bibinfo{author}{\bibfnamefont{M.}~\bibnamefont{Della~Morte}},
  \bibinfo{author}{\bibfnamefont{A.}~\bibnamefont{Francis}},
  \bibinfo{author}{\bibfnamefont{V.}~\bibnamefont{G{\"u}lpers}},
  \bibinfo{author}{\bibfnamefont{G.}~\bibnamefont{Herdo{\'i}za}},
  \bibinfo{author}{\bibfnamefont{G.}~\bibnamefont{von Hippel}},
  \bibinfo{author}{\bibfnamefont{H.}~\bibnamefont{Horch}},
  \bibinfo{author}{\bibfnamefont{B.}~\bibnamefont{J{\"a}ger}},
  \bibinfo{author}{\bibfnamefont{H.~B.} \bibnamefont{Meyer}},
  \bibinfo{author}{\bibfnamefont{A.}~\bibnamefont{Nyffeler}}, \bibnamefont{and}
  \bibinfo{author}{\bibfnamefont{H.}~\bibnamefont{Wittig}},
  \bibinfo{journal}{JHEP} \textbf{\bibinfo{volume}{10}}, \bibinfo{pages}{020}
  (\bibinfo{year}{2017}), \eprint{1705.01775}.

\bibitem[{\citenamefont{Della~Morte et~al.}(2018)}]{DellaMorte:2017khn}
\bibinfo{author}{\bibfnamefont{M.}~\bibnamefont{Della~Morte}}
  \bibnamefont{et~al.}, \bibinfo{journal}{EPJ Web Conf.}
  \textbf{\bibinfo{volume}{175}}, \bibinfo{pages}{06031}
  (\bibinfo{year}{2018}), \eprint{1710.10072}.

\bibitem[{\citenamefont{Feng et~al.}(2011)\citenamefont{Feng, Jansen,
  Petschlies, and Renner}}]{Feng:2011zk}
\bibinfo{author}{\bibfnamefont{X.}~\bibnamefont{Feng}},
  \bibinfo{author}{\bibfnamefont{K.}~\bibnamefont{Jansen}},
  \bibinfo{author}{\bibfnamefont{M.}~\bibnamefont{Petschlies}},
  \bibnamefont{and} \bibinfo{author}{\bibfnamefont{D.~B.}
  \bibnamefont{Renner}}, \bibinfo{journal}{Phys.Rev.Lett.}
  \textbf{\bibinfo{volume}{107}}, \bibinfo{pages}{081802}
  (\bibinfo{year}{2011}), \eprint{1103.4818}.

\bibitem[{\citenamefont{L{\"u}scher and Schaefer}(2013)}]{Luscher:2012av}
\bibinfo{author}{\bibfnamefont{M.}~\bibnamefont{L{\"u}scher}} \bibnamefont{and}
  \bibinfo{author}{\bibfnamefont{S.}~\bibnamefont{Schaefer}},
  \bibinfo{journal}{Comput.Phys.Commun.} \textbf{\bibinfo{volume}{184}},
  \bibinfo{pages}{519} (\bibinfo{year}{2013}), \eprint{1206.2809}.

\bibitem[{\citenamefont{Bulava and Schaefer}(2013)}]{Bulava:2013cta}
\bibinfo{author}{\bibfnamefont{J.}~\bibnamefont{Bulava}} \bibnamefont{and}
  \bibinfo{author}{\bibfnamefont{S.}~\bibnamefont{Schaefer}},
  \bibinfo{journal}{Nucl.Phys.} \textbf{\bibinfo{volume}{B874}},
  \bibinfo{pages}{188} (\bibinfo{year}{2013}), \eprint{1304.7093}.

\bibitem[{\citenamefont{Bruno et~al.}(2017)\citenamefont{Bruno, Korzec, and
  Schaefer}}]{Bruno:2016plf}
\bibinfo{author}{\bibfnamefont{M.}~\bibnamefont{Bruno}},
  \bibinfo{author}{\bibfnamefont{T.}~\bibnamefont{Korzec}}, \bibnamefont{and}
  \bibinfo{author}{\bibfnamefont{S.}~\bibnamefont{Schaefer}},
  \bibinfo{journal}{Phys. Rev.} \textbf{\bibinfo{volume}{D95}},
  \bibinfo{pages}{074504} (\bibinfo{year}{2017}), \eprint{1608.08900}.

\bibitem[{\citenamefont{L{\"u}scher and Schaefer}(2011)}]{Luscher:2011kk}
\bibinfo{author}{\bibfnamefont{M.}~\bibnamefont{L{\"u}scher}} \bibnamefont{and}
  \bibinfo{author}{\bibfnamefont{S.}~\bibnamefont{Schaefer}},
  \bibinfo{journal}{JHEP} \textbf{\bibinfo{volume}{07}}, \bibinfo{pages}{036}
  (\bibinfo{year}{2011}), \eprint{1105.4749}.

\bibitem[{\citenamefont{Wilcox}(2000)}]{Wilcox:1999ab}
\bibinfo{author}{\bibfnamefont{W.}~\bibnamefont{Wilcox}}, in
  \emph{\bibinfo{booktitle}{Numerical Challenges in Lattice Quantum
  Chromodynamics}}, edited by
  \bibinfo{editor}{\bibfnamefont{A.}~\bibnamefont{Frommer}},
  \bibinfo{editor}{\bibfnamefont{T.}~\bibnamefont{Lippert}},
  \bibinfo{editor}{\bibfnamefont{B.}~\bibnamefont{Medeke}}, \bibnamefont{and}
  \bibinfo{editor}{\bibfnamefont{K.}~\bibnamefont{Schilling}}
  (\bibinfo{publisher}{Springer Berlin Heidelberg}, \bibinfo{year}{2000}),
  vol.~\bibinfo{volume}{15} of \emph{\bibinfo{series}{Lecture Notes in
  Computational Science and Engineering}}, pp. \bibinfo{pages}{127--141}, ISBN
  \bibinfo{isbn}{978-3-540-67732-1}, \eprint{hep-lat/9911013}.

\bibitem[{\citenamefont{Foley et~al.}(2005)\citenamefont{Foley, Jimmy~Juge,
  O'Cais, Peardon, Ryan et~al.}}]{Foley:2005ac}
\bibinfo{author}{\bibfnamefont{J.}~\bibnamefont{Foley}},
  \bibinfo{author}{\bibfnamefont{K.}~\bibnamefont{Jimmy~Juge}},
  \bibinfo{author}{\bibfnamefont{A.}~\bibnamefont{O'Cais}},
  \bibinfo{author}{\bibfnamefont{M.}~\bibnamefont{Peardon}},
  \bibinfo{author}{\bibfnamefont{S.~M.} \bibnamefont{Ryan}},
  \bibnamefont{et~al.}, \bibinfo{journal}{Comput.Phys.Commun.}
  \textbf{\bibinfo{volume}{172}}, \bibinfo{pages}{145} (\bibinfo{year}{2005}),
  \eprint{hep-lat/0505023}.

\bibitem[{\citenamefont{Andersen et~al.}(2019)\citenamefont{Andersen, Bulava,
  H{\"o}rz, and Morningstar}}]{Andersen:2018mau}
\bibinfo{author}{\bibfnamefont{C.}~\bibnamefont{Andersen}},
  \bibinfo{author}{\bibfnamefont{J.}~\bibnamefont{Bulava}},
  \bibinfo{author}{\bibfnamefont{B.}~\bibnamefont{H{\"o}rz}}, \bibnamefont{and}
  \bibinfo{author}{\bibfnamefont{C.}~\bibnamefont{Morningstar}},
  \bibinfo{journal}{Nucl. Phys.} \textbf{\bibinfo{volume}{B939}},
  \bibinfo{pages}{145} (\bibinfo{year}{2019}), \eprint{1808.05007}.

\bibitem[{\citenamefont{Stathopoulos et~al.}(2013)\citenamefont{Stathopoulos,
  Laeuchli, and Orginos}}]{Stathopoulos:2013aci}
\bibinfo{author}{\bibfnamefont{A.}~\bibnamefont{Stathopoulos}},
  \bibinfo{author}{\bibfnamefont{J.}~\bibnamefont{Laeuchli}}, \bibnamefont{and}
  \bibinfo{author}{\bibfnamefont{K.}~\bibnamefont{Orginos}}
  (\bibinfo{year}{2013}), \eprint{1302.4018}.

\bibitem[{\citenamefont{Djukanovic et~al.}(2019)\citenamefont{Djukanovic,
  Ottnad, Wilhelm, and Wittig}}]{Djukanovic:2019jtp}
\bibinfo{author}{\bibfnamefont{D.}~\bibnamefont{Djukanovic}},
  \bibinfo{author}{\bibfnamefont{K.}~\bibnamefont{Ottnad}},
  \bibinfo{author}{\bibfnamefont{J.}~\bibnamefont{Wilhelm}}, \bibnamefont{and}
  \bibinfo{author}{\bibfnamefont{H.}~\bibnamefont{Wittig}}
  (\bibinfo{year}{2019}), \eprint{1903.12566}.

\bibitem[{\citenamefont{L{\"u}scher et~al.}(1996)\citenamefont{L{\"u}scher,
  Sint, Sommer, and Weisz}}]{Luscher:1996sc}
\bibinfo{author}{\bibfnamefont{M.}~\bibnamefont{L{\"u}scher}},
  \bibinfo{author}{\bibfnamefont{S.}~\bibnamefont{Sint}},
  \bibinfo{author}{\bibfnamefont{R.}~\bibnamefont{Sommer}}, \bibnamefont{and}
  \bibinfo{author}{\bibfnamefont{P.}~\bibnamefont{Weisz}},
  \bibinfo{journal}{Nucl. Phys.} \textbf{\bibinfo{volume}{B478}},
  \bibinfo{pages}{365} (\bibinfo{year}{1996}), \eprint{hep-lat/9605038}.

\bibitem[{\citenamefont{Bhattacharya et~al.}(2006)\citenamefont{Bhattacharya,
  Gupta, Lee, Sharpe, and Wu}}]{Bhattacharya:2005rb}
\bibinfo{author}{\bibfnamefont{T.}~\bibnamefont{Bhattacharya}},
  \bibinfo{author}{\bibfnamefont{R.}~\bibnamefont{Gupta}},
  \bibinfo{author}{\bibfnamefont{W.}~\bibnamefont{Lee}},
  \bibinfo{author}{\bibfnamefont{S.~R.} \bibnamefont{Sharpe}},
  \bibnamefont{and} \bibinfo{author}{\bibfnamefont{J.~M.~S.} \bibnamefont{Wu}},
  \bibinfo{journal}{Phys. Rev.} \textbf{\bibinfo{volume}{D73}},
  \bibinfo{pages}{034504} (\bibinfo{year}{2006}), \eprint{hep-lat/0511014}.

\bibitem[{\citenamefont{G{\'e}rardin
  et~al.}(2019{\natexlab{a}})\citenamefont{G{\'e}rardin, Harris, and
  Meyer}}]{Gerardin:2018kpy}
\bibinfo{author}{\bibfnamefont{A.}~\bibnamefont{G{\'e}rardin}},
  \bibinfo{author}{\bibfnamefont{T.}~\bibnamefont{Harris}}, \bibnamefont{and}
  \bibinfo{author}{\bibfnamefont{H.~B.} \bibnamefont{Meyer}},
  \bibinfo{journal}{Phys. Rev.} \textbf{\bibinfo{volume}{D99}},
  \bibinfo{pages}{014519} (\bibinfo{year}{2019}{\natexlab{a}}),
  \eprint{1811.08209}.

\bibitem[{\citenamefont{Dalla~Brida et~al.}(2019)\citenamefont{Dalla~Brida,
  Korzec, Sint, and Vilaseca}}]{DallaBrida:2018tpn}
\bibinfo{author}{\bibfnamefont{M.}~\bibnamefont{Dalla~Brida}},
  \bibinfo{author}{\bibfnamefont{T.}~\bibnamefont{Korzec}},
  \bibinfo{author}{\bibfnamefont{S.}~\bibnamefont{Sint}}, \bibnamefont{and}
  \bibinfo{author}{\bibfnamefont{P.}~\bibnamefont{Vilaseca}},
  \bibinfo{journal}{Eur. Phys. J.} \textbf{\bibinfo{volume}{C79}},
  \bibinfo{pages}{23} (\bibinfo{year}{2019}), \eprint{1808.09236}.

\bibitem[{\citenamefont{Harris and Meyer}(2015)}]{Harris:2015vfa}
\bibinfo{author}{\bibfnamefont{T.}~\bibnamefont{Harris}} \bibnamefont{and}
  \bibinfo{author}{\bibfnamefont{H.~B.} \bibnamefont{Meyer}},
  \bibinfo{journal}{Phys. Rev.} \textbf{\bibinfo{volume}{D92}},
  \bibinfo{pages}{114503} (\bibinfo{year}{2015}), \eprint{1506.05248}.

\bibitem[{\citenamefont{Lehner}(2016)}]{CLehnerBounding}
\bibinfo{author}{\bibfnamefont{C.}~\bibnamefont{Lehner}},
  \bibinfo{journal}{RBRC Workshop on Lattice Gauge Theories}
  (\bibinfo{year}{2016}).

\bibitem[{\citenamefont{Borsanyi et~al.}(2018)}]{Borsanyi:2017zdw}
\bibinfo{author}{\bibfnamefont{S.}~\bibnamefont{Borsanyi}} \bibnamefont{et~al.}
  (\bibinfo{collaboration}{Budapest-Marseille-Wuppertal}),
  \bibinfo{journal}{Phys. Rev. Lett.} \textbf{\bibinfo{volume}{121}},
  \bibinfo{pages}{022002} (\bibinfo{year}{2018}), \eprint{1711.04980}.

\bibitem[{\citenamefont{Blum et~al.}(2018)\citenamefont{Blum, Boyle,
  G{\"u}lpers, Izubuchi, Jin, Jung, J{\"u}ttner, Lehner, Portelli, and
  Tsang}}]{Blum:2018mom}
\bibinfo{author}{\bibfnamefont{T.}~\bibnamefont{Blum}},
  \bibinfo{author}{\bibfnamefont{P.~A.} \bibnamefont{Boyle}},
  \bibinfo{author}{\bibfnamefont{V.}~\bibnamefont{G{\"u}lpers}},
  \bibinfo{author}{\bibfnamefont{T.}~\bibnamefont{Izubuchi}},
  \bibinfo{author}{\bibfnamefont{L.}~\bibnamefont{Jin}},
  \bibinfo{author}{\bibfnamefont{C.}~\bibnamefont{Jung}},
  \bibinfo{author}{\bibfnamefont{A.}~\bibnamefont{J{\"u}ttner}},
  \bibinfo{author}{\bibfnamefont{C.}~\bibnamefont{Lehner}},
  \bibinfo{author}{\bibfnamefont{A.}~\bibnamefont{Portelli}}, \bibnamefont{and}
  \bibinfo{author}{\bibfnamefont{J.~T.} \bibnamefont{Tsang}}
  (\bibinfo{collaboration}{RBC, UKQCD}), \bibinfo{journal}{Phys. Rev. Lett.}
  \textbf{\bibinfo{volume}{121}}, \bibinfo{pages}{022003}
  (\bibinfo{year}{2018}), \eprint{1801.07224}.

\bibitem[{\citenamefont{Aubin and Blum}(2007)}]{Aubin:2006xv}
\bibinfo{author}{\bibfnamefont{C.}~\bibnamefont{Aubin}} \bibnamefont{and}
  \bibinfo{author}{\bibfnamefont{T.}~\bibnamefont{Blum}},
  \bibinfo{journal}{Phys.Rev.} \textbf{\bibinfo{volume}{D75}},
  \bibinfo{pages}{114502} (\bibinfo{year}{2007}), \eprint{hep-lat/0608011}.

\bibitem[{\citenamefont{Francis et~al.}(2013)\citenamefont{Francis, J{\"a}ger,
  Meyer, and Wittig}}]{Francis:2013fzp}
\bibinfo{author}{\bibfnamefont{A.}~\bibnamefont{Francis}},
  \bibinfo{author}{\bibfnamefont{B.}~\bibnamefont{J{\"a}ger}},
  \bibinfo{author}{\bibfnamefont{H.~B.} \bibnamefont{Meyer}}, \bibnamefont{and}
  \bibinfo{author}{\bibfnamefont{H.}~\bibnamefont{Wittig}},
  \bibinfo{journal}{Phys.Rev.} \textbf{\bibinfo{volume}{D88}},
  \bibinfo{pages}{054502} (\bibinfo{year}{2013}), \eprint{1306.2532}.

\bibitem[{\citenamefont{Aubin et~al.}(2016)\citenamefont{Aubin, Blum, Chau,
  Golterman, Peris, and Tu}}]{Aubin:2015rzx}
\bibinfo{author}{\bibfnamefont{C.}~\bibnamefont{Aubin}},
  \bibinfo{author}{\bibfnamefont{T.}~\bibnamefont{Blum}},
  \bibinfo{author}{\bibfnamefont{P.}~\bibnamefont{Chau}},
  \bibinfo{author}{\bibfnamefont{M.}~\bibnamefont{Golterman}},
  \bibinfo{author}{\bibfnamefont{S.}~\bibnamefont{Peris}}, \bibnamefont{and}
  \bibinfo{author}{\bibfnamefont{C.}~\bibnamefont{Tu}}, \bibinfo{journal}{Phys.
  Rev.} \textbf{\bibinfo{volume}{D93}}, \bibinfo{pages}{054508}
  (\bibinfo{year}{2016}), \eprint{1512.07555}.

\bibitem[{\citenamefont{Gounaris and Sakurai}(1968)}]{Gounaris:1968mw}
\bibinfo{author}{\bibfnamefont{G.}~\bibnamefont{Gounaris}} \bibnamefont{and}
  \bibinfo{author}{\bibfnamefont{J.}~\bibnamefont{Sakurai}},
  \bibinfo{journal}{Phys.Rev.Lett.} \textbf{\bibinfo{volume}{21}},
  \bibinfo{pages}{244} (\bibinfo{year}{1968}).

\bibitem[{\citenamefont{Feng et~al.}(2015)\citenamefont{Feng, Aoki, Hashimoto,
  and Kaneko}}]{Feng:2014gba}
\bibinfo{author}{\bibfnamefont{X.}~\bibnamefont{Feng}},
  \bibinfo{author}{\bibfnamefont{S.}~\bibnamefont{Aoki}},
  \bibinfo{author}{\bibfnamefont{S.}~\bibnamefont{Hashimoto}},
  \bibnamefont{and} \bibinfo{author}{\bibfnamefont{T.}~\bibnamefont{Kaneko}},
  \bibinfo{journal}{Phys. Rev.} \textbf{\bibinfo{volume}{D91}},
  \bibinfo{pages}{054504} (\bibinfo{year}{2015}), \eprint{1412.6319}.

\bibitem[{\citenamefont{Golterman et~al.}(2017)\citenamefont{Golterman,
  Maltman, and Peris}}]{Golterman:2017njs}
\bibinfo{author}{\bibfnamefont{M.}~\bibnamefont{Golterman}},
  \bibinfo{author}{\bibfnamefont{K.}~\bibnamefont{Maltman}}, \bibnamefont{and}
  \bibinfo{author}{\bibfnamefont{S.}~\bibnamefont{Peris}},
  \bibinfo{journal}{Phys. Rev.} \textbf{\bibinfo{volume}{D95}},
  \bibinfo{pages}{074509} (\bibinfo{year}{2017}), \eprint{1701.08685}.

\bibitem[{\citenamefont{Golowich and Kambor}(1995)}]{Golowich:1995kd}
\bibinfo{author}{\bibfnamefont{E.}~\bibnamefont{Golowich}} \bibnamefont{and}
  \bibinfo{author}{\bibfnamefont{J.}~\bibnamefont{Kambor}},
  \bibinfo{journal}{Nucl. Phys.} \textbf{\bibinfo{volume}{B447}},
  \bibinfo{pages}{373} (\bibinfo{year}{1995}), \eprint{hep-ph/9501318}.

\bibitem[{\citenamefont{Amoros et~al.}(2000)\citenamefont{Amoros, Bijnens, and
  Talavera}}]{Amoros:1999dp}
\bibinfo{author}{\bibfnamefont{G.}~\bibnamefont{Amoros}},
  \bibinfo{author}{\bibfnamefont{J.}~\bibnamefont{Bijnens}}, \bibnamefont{and}
  \bibinfo{author}{\bibfnamefont{P.}~\bibnamefont{Talavera}},
  \bibinfo{journal}{Nucl. Phys.} \textbf{\bibinfo{volume}{B568}},
  \bibinfo{pages}{319} (\bibinfo{year}{2000}), \eprint{hep-ph/9907264}.

\bibitem[{\citenamefont{Colangelo et~al.}(2005)\citenamefont{Colangelo, Durr,
  and Haefeli}}]{Colangelo:2005gd}
\bibinfo{author}{\bibfnamefont{G.}~\bibnamefont{Colangelo}},
  \bibinfo{author}{\bibfnamefont{S.}~\bibnamefont{Durr}}, \bibnamefont{and}
  \bibinfo{author}{\bibfnamefont{C.}~\bibnamefont{Haefeli}},
  \bibinfo{journal}{Nucl.Phys.} \textbf{\bibinfo{volume}{B721}},
  \bibinfo{pages}{136} (\bibinfo{year}{2005}), \eprint{hep-lat/0503014}.

\bibitem[{\citenamefont{Alexandrou et~al.}(2017)\citenamefont{Alexandrou,
  Leskovec, Meinel, Negele, Paul, Petschlies, Pochinsky, Rendon, and
  Syritsyn}}]{Alexandrou:2017mpi}
\bibinfo{author}{\bibfnamefont{C.}~\bibnamefont{Alexandrou}},
  \bibinfo{author}{\bibfnamefont{L.}~\bibnamefont{Leskovec}},
  \bibinfo{author}{\bibfnamefont{S.}~\bibnamefont{Meinel}},
  \bibinfo{author}{\bibfnamefont{J.}~\bibnamefont{Negele}},
  \bibinfo{author}{\bibfnamefont{S.}~\bibnamefont{Paul}},
  \bibinfo{author}{\bibfnamefont{M.}~\bibnamefont{Petschlies}},
  \bibinfo{author}{\bibfnamefont{A.}~\bibnamefont{Pochinsky}},
  \bibinfo{author}{\bibfnamefont{G.}~\bibnamefont{Rendon}}, \bibnamefont{and}
  \bibinfo{author}{\bibfnamefont{S.}~\bibnamefont{Syritsyn}},
  \bibinfo{journal}{Phys. Rev.} \textbf{\bibinfo{volume}{D96}},
  \bibinfo{pages}{034525} (\bibinfo{year}{2017}), \eprint{1704.05439}.

\bibitem[{\citenamefont{Fu and Wang}(2016)}]{Fu:2016itp}
\bibinfo{author}{\bibfnamefont{Z.}~\bibnamefont{Fu}} \bibnamefont{and}
  \bibinfo{author}{\bibfnamefont{L.}~\bibnamefont{Wang}},
  \bibinfo{journal}{Phys. Rev.} \textbf{\bibinfo{volume}{D94}},
  \bibinfo{pages}{034505} (\bibinfo{year}{2016}), \eprint{1608.07478}.

\bibitem[{\citenamefont{Guo et~al.}(2016)\citenamefont{Guo, Alexandru, Molina,
  and {D\"{o}ring}}}]{Guo:2016zos}
\bibinfo{author}{\bibfnamefont{D.}~\bibnamefont{Guo}},
  \bibinfo{author}{\bibfnamefont{A.}~\bibnamefont{Alexandru}},
  \bibinfo{author}{\bibfnamefont{R.}~\bibnamefont{Molina}}, \bibnamefont{and}
  \bibinfo{author}{\bibfnamefont{M.}~\bibnamefont{{D\"{o}ring}}},
  \bibinfo{journal}{Phys. Rev.} \textbf{\bibinfo{volume}{D94}},
  \bibinfo{pages}{034501} (\bibinfo{year}{2016}), \eprint{1605.03993}.

\bibitem[{\citenamefont{Wilson et~al.}(2015)\citenamefont{Wilson, Briceno,
  Dudek, Edwards, and Thomas}}]{Wilson:2015dqa}
\bibinfo{author}{\bibfnamefont{D.~J.} \bibnamefont{Wilson}},
  \bibinfo{author}{\bibfnamefont{R.~A.} \bibnamefont{Briceno}},
  \bibinfo{author}{\bibfnamefont{J.~J.} \bibnamefont{Dudek}},
  \bibinfo{author}{\bibfnamefont{R.~G.} \bibnamefont{Edwards}},
  \bibnamefont{and} \bibinfo{author}{\bibfnamefont{C.~E.}
  \bibnamefont{Thomas}}, \bibinfo{journal}{Phys. Rev.}
  \textbf{\bibinfo{volume}{D92}}, \bibinfo{pages}{094502}
  (\bibinfo{year}{2015}), \eprint{1507.02599}.

\bibitem[{\citenamefont{Bali et~al.}(2016{\natexlab{b}})\citenamefont{Bali,
  Collins, Cox, Donald, G{\"o}ckeler, Lang, and Sch{\"a}fer}}]{Bali:2015gji}
\bibinfo{author}{\bibfnamefont{G.~S.} \bibnamefont{Bali}},
  \bibinfo{author}{\bibfnamefont{S.}~\bibnamefont{Collins}},
  \bibinfo{author}{\bibfnamefont{A.}~\bibnamefont{Cox}},
  \bibinfo{author}{\bibfnamefont{G.}~\bibnamefont{Donald}},
  \bibinfo{author}{\bibfnamefont{M.}~\bibnamefont{G{\"o}ckeler}},
  \bibinfo{author}{\bibfnamefont{C.~B.} \bibnamefont{Lang}}, \bibnamefont{and}
  \bibinfo{author}{\bibfnamefont{A.}~\bibnamefont{Sch{\"a}fer}}
  (\bibinfo{collaboration}{RQCD}), \bibinfo{journal}{Phys. Rev.}
  \textbf{\bibinfo{volume}{D93}}, \bibinfo{pages}{054509}
  (\bibinfo{year}{2016}{\natexlab{b}}), \eprint{1512.08678}.

\bibitem[{\citenamefont{Giusti et~al.}(2019)\citenamefont{Giusti, Lubicz,
  Martinelli, Sanfilippo, and Simula}}]{Giusti:2019xct}
\bibinfo{author}{\bibfnamefont{D.}~\bibnamefont{Giusti}},
  \bibinfo{author}{\bibfnamefont{V.}~\bibnamefont{Lubicz}},
  \bibinfo{author}{\bibfnamefont{G.}~\bibnamefont{Martinelli}},
  \bibinfo{author}{\bibfnamefont{F.}~\bibnamefont{Sanfilippo}},
  \bibnamefont{and} \bibinfo{author}{\bibfnamefont{S.}~\bibnamefont{Simula}}
  (\bibinfo{year}{2019}), \eprint{1901.10462}.

\bibitem[{\citenamefont{Keshavarzi et~al.}(2018)\citenamefont{Keshavarzi,
  Nomura, and Teubner}}]{Keshavarzi:2018mgv}
\bibinfo{author}{\bibfnamefont{A.}~\bibnamefont{Keshavarzi}},
  \bibinfo{author}{\bibfnamefont{D.}~\bibnamefont{Nomura}}, \bibnamefont{and}
  \bibinfo{author}{\bibfnamefont{T.}~\bibnamefont{Teubner}},
  \bibinfo{journal}{Phys. Rev.} \textbf{\bibinfo{volume}{D97}},
  \bibinfo{pages}{114025} (\bibinfo{year}{2018}), \eprint{1802.02995}.

\bibitem[{\citenamefont{Chakraborty et~al.}(2014)\citenamefont{Chakraborty,
  Davies, Donald, Dowdall, Koponen, Lepage, and Teubner}}]{Chakraborty:2014mwa}
\bibinfo{author}{\bibfnamefont{B.}~\bibnamefont{Chakraborty}},
  \bibinfo{author}{\bibfnamefont{C.~T.~H.} \bibnamefont{Davies}},
  \bibinfo{author}{\bibfnamefont{G.~C.} \bibnamefont{Donald}},
  \bibinfo{author}{\bibfnamefont{R.~J.} \bibnamefont{Dowdall}},
  \bibinfo{author}{\bibfnamefont{J.}~\bibnamefont{Koponen}},
  \bibinfo{author}{\bibfnamefont{G.~P.} \bibnamefont{Lepage}},
  \bibnamefont{and} \bibinfo{author}{\bibfnamefont{T.}~\bibnamefont{Teubner}}
  (\bibinfo{collaboration}{HPQCD}), \bibinfo{journal}{Phys. Rev.}
  \textbf{\bibinfo{volume}{D89}}, \bibinfo{pages}{114501}
  (\bibinfo{year}{2014}), \eprint{1403.1778}.

\bibitem[{\citenamefont{Chakraborty et~al.}(2018)}]{Chakraborty:2017tqp}
\bibinfo{author}{\bibfnamefont{B.}~\bibnamefont{Chakraborty}}
  \bibnamefont{et~al.} (\bibinfo{collaboration}{Fermilab Lattice,
  LATTICE-HPQCD, MILC}), \bibinfo{journal}{Phys. Rev. Lett.}
  \textbf{\bibinfo{volume}{120}}, \bibinfo{pages}{152001}
  (\bibinfo{year}{2018}), \eprint{1710.11212}.

\bibitem[{\citenamefont{Davies et~al.}(2019)}]{Davies:2019efs}
\bibinfo{author}{\bibfnamefont{C.~T.~H.} \bibnamefont{Davies}}
  \bibnamefont{et~al.} (\bibinfo{collaboration}{Fermilab Lattice,
  LATTICE-HPQCD, MILC}) (\bibinfo{year}{2019}), \eprint{1902.04223}.

\bibitem[{\citenamefont{Shintani and Kuramashi}(2019)}]{Shintani:2019wai}
\bibinfo{author}{\bibfnamefont{E.}~\bibnamefont{Shintani}} \bibnamefont{and}
  \bibinfo{author}{\bibfnamefont{Y.}~\bibnamefont{Kuramashi}}
  (\bibinfo{year}{2019}), \eprint{1902.00885}.

\bibitem[{\citenamefont{Giusti et~al.}(2017)\citenamefont{Giusti, Lubicz,
  Martinelli, Sanfilippo, and Simula}}]{Giusti:2017jof}
\bibinfo{author}{\bibfnamefont{D.}~\bibnamefont{Giusti}},
  \bibinfo{author}{\bibfnamefont{V.}~\bibnamefont{Lubicz}},
  \bibinfo{author}{\bibfnamefont{G.}~\bibnamefont{Martinelli}},
  \bibinfo{author}{\bibfnamefont{F.}~\bibnamefont{Sanfilippo}},
  \bibnamefont{and} \bibinfo{author}{\bibfnamefont{S.}~\bibnamefont{Simula}},
  \bibinfo{journal}{JHEP} \textbf{\bibinfo{volume}{10}}, \bibinfo{pages}{157}
  (\bibinfo{year}{2017}), \eprint{1707.03019}.

\bibitem[{\citenamefont{Giusti et~al.}(2018)\citenamefont{Giusti, Sanfilippo,
  and Simula}}]{Giusti:2018mdh}
\bibinfo{author}{\bibfnamefont{D.}~\bibnamefont{Giusti}},
  \bibinfo{author}{\bibfnamefont{F.}~\bibnamefont{Sanfilippo}},
  \bibnamefont{and} \bibinfo{author}{\bibfnamefont{S.}~\bibnamefont{Simula}},
  \bibinfo{journal}{Phys. Rev.} \textbf{\bibinfo{volume}{D98}},
  \bibinfo{pages}{114504} (\bibinfo{year}{2018}), \eprint{1808.00887}.

\bibitem[{\citenamefont{Boyle et~al.}(2017)\citenamefont{Boyle, Chuvelev,
  Cossu, Kelly, Lehner, and Meadows}}]{Boyle:2017xcy}
\bibinfo{author}{\bibfnamefont{P.}~\bibnamefont{Boyle}},
  \bibinfo{author}{\bibfnamefont{M.}~\bibnamefont{Chuvelev}},
  \bibinfo{author}{\bibfnamefont{G.}~\bibnamefont{Cossu}},
  \bibinfo{author}{\bibfnamefont{C.}~\bibnamefont{Kelly}},
  \bibinfo{author}{\bibfnamefont{C.}~\bibnamefont{Lehner}}, \bibnamefont{and}
  \bibinfo{author}{\bibfnamefont{L.}~\bibnamefont{Meadows}}
  (\bibinfo{year}{2017}), \eprint{1711.04883}.

\bibitem[{\citenamefont{Edwards and Joo}(2005)}]{Edwards:2004sx}
\bibinfo{author}{\bibfnamefont{R.~G.} \bibnamefont{Edwards}} \bibnamefont{and}
  \bibinfo{author}{\bibfnamefont{B.}~\bibnamefont{Joo}}
  (\bibinfo{collaboration}{SciDAC, LHPC, UKQCD}), \bibinfo{journal}{Nucl. Phys.
  Proc. Suppl.} \textbf{\bibinfo{volume}{140}}, \bibinfo{pages}{832}
  (\bibinfo{year}{2005}), \eprint{hep-lat/0409003}.

\bibitem[{\citenamefont{G{\'e}rardin
  et~al.}(2019{\natexlab{b}})\citenamefont{G{\'e}rardin, Meyer, and
  Nyffeler}}]{Gerardin:2019vio}
\bibinfo{author}{\bibfnamefont{A.}~\bibnamefont{G{\'e}rardin}},
  \bibinfo{author}{\bibfnamefont{H.~B.} \bibnamefont{Meyer}}, \bibnamefont{and}
  \bibinfo{author}{\bibfnamefont{A.}~\bibnamefont{Nyffeler}}
  (\bibinfo{year}{2019}{\natexlab{b}}), \eprint{1903.09471}.

\end{thebibliography}

\end{document}